\documentclass{ar2e}

\def\del{\partial}
\def\F{{\cal F}}
\def\A{{\cal A}}

\def\me#1#2#3{\langle #1 \vert #2 \vert #3 \rangle}

\def\beq{\begin{equation}}
\def\eeq{\end{equation}}
\def\beqn{\begin{eqnarray}}
\def\eeqn{\end{eqnarray}}
\usepackage{amssymb}
\usepackage{amsmath}
\usepackage{graphics}
\renewcommand{\numberline}[1]{}
\usepackage{graphicx}
\usepackage{dcolumn}
\usepackage{bm}

\begin{document}

\input epsf.def
\input psfig.sty

\jname{Annu. Rev. Condensed Matter}
\jyear{2010}
\jvol{1}
\ARinfo{00}

\title{Three-Dimensional Topological Insulators}
\markboth{Hasan \& Moore}{Three-Dimensional Topological Insulators}

\author{M. Zahid Hasan
\affiliation{Joseph Henry Laboratory, Department of Physics, Princeton University, Princeton, NJ 08544, and Princeton Center for Complex Materials, PRISM, Princeton University, Princeton, NJ 08544}
Joel E. Moore
\affiliation{Department of Physics, University of California, Berkeley, California 94720, and Materials Sciences Division, Lawrence Berkeley National Laboratory, Berkeley, California 94720}
}
\begin{abstract}
Topological insulators in three dimensions are nonmagnetic insulators that possess metallic surface states as a consequence of the nontrivial topology of electronic wavefunctions in the bulk of the material.  They are the first known examples of topological order in bulk solids.  We review the basic phenomena and experimental history, starting with the observation of topological insulator behavior in Bi$_x$Sb$_{1-x}$ by spin- and angle-resolved photoemission spectroscopy and continuing through measurements on other materials and by other probes.  A self-contained introduction to the single-particle theory is then given, followed by the many-particle definition of a topological insulator as a material with quantized magnetoelectric polarizability.  The last section reviews recent work on strongly correlated topological insulators and new effects that arise from the proximity effect between a topological insulator and a superconductor.  While this article is not intended to be a comprehensive review of what is already a rather large field, we hope that it will serve as a useful introduction, summary of recent progress, and guideline to future directions.
\end{abstract}

\maketitle

\section{INTRODUCTION}

Condensed matter physics starts at microscopic scales with a clear notion of spacetime distance defined by the Minkowski metric of special relativity.  It has been understood since the discovery of the integer quantum Hall effect (IQHE) that some phases of matter, known as topological phases, reflect the ``emergence'' of a different type of spacetime.  The notion of a metric disappears entirely, and macroscopic physics is controlled by properties that are insensitive to spacetime distance and described by the branch of mathematics known as topology.  The emergence of a topological description reflects a type of order in condensed matter physics that is quite different from conventional orders described in terms of symmetry breaking.  This review discusses recent experimental and theoretical progress on three-dimensional topological insulators, the first topological phase of bulk solids.

One distinguishing feature of topological phases is that they are typically insulating or ``gapped'' in bulk but possess gapless edge or surface states as a result of the topology.  A related property is that the topological order can lead to measurable physical properties with extremely precise quantization as a result of the insensitivity of topology to local perturbations such as defects and thermal fluctuations; the first example was the Hall (i.e., transverse) conductance in the IQHE, which can be quantized to an integer multiple of $e^2/h$ with an accuracy of one part per billion.  However, topological order until recently was only observed in two-dimensional electron gases under extreme conditions of low temperature and high magnetic field.

The current excitement in this field results from the experimental discovery that there are topological phases in fairly ordinary materials that occur in three dimensions and zero magnetic field~\cite{hsieh,hasankane,moorenature}.  The existence of three-dimensional (3D)`` topological insulators''~\cite{moore&balents-2006} was predicted by deriving topological invariants for a general time-reversal-invariant band insulator that built on previous work~\cite{kane&mele2-2005} on a new topological invariant of time-reversal-invariant band insulators in 2D.  Topological invariants are mathematical quantities, in this case essentially integrals over a band structure, that are stable to continuous changes of parameters.  Simpler forms of these invariants in inversion-symmetric materials were found independently~\cite{fu&kane&mele-2007} and used to predict the first topological insulator to be discovered, Bi$_x$Sb$_{1-x}$~\cite{fu&kane2-2006}.  The 2D topological invariant underlies the ``quantum spin Hall effect'' observed in quantum wells derived from HgTe~\cite{zhangscience1,molenkampscience}.  The general term ``topological insulators'' was coined to indicate that both the 2D and 3D phases are topological in the same sense as the IQHE, with topologically protected edge or surface states that result from spin-orbit coupling rather than a magnetic field.

The method used to discover all 3D topological insulators known so far is angle-resolved photoemission spectroscopy (ARPES), which probes directly the unique metallic surface state and consequently the topological invariants.  The accessibility of 3D topological insulators to a variety of experimental probes means that many other types of data are now available.  Topological insulators in several materials families have been discovered to date, and theoretical work has connected topological insulators to many other areas of condensed matter physics.  As both technical~\cite{hasankane} and non-technical~\cite{qizhang,moorenature} reviews are available that introduce three-dimensional topological insulators following the historical development from the 2D case, and there is a published review specifically on the 2D case~\cite{molenkampscience,qshereview}, we seek here to give a reasonably self-contained treatment of the 3D case, which is the focus of most current interest for both practical and fundamental reasons.

Given length limitations and the existing review of the literature as of early 2010~\cite{hasankane}, we will seek here to give an overview of basic facts and the essential theoretical background, then concentrate on recent developments.  We first review the basic facts and experimental discoveries in Section II.  This section includes a discussion of which materials are currently believed to be topological insulators and how these materials are characterized by various techniques.  Our discussion of the theory of topological insulators and how they connect to several long-standing projects of theoretical physics, makes up Section III.  Discussing even a few aspects in depth means that there are omissions in other areas; in particular, we will have little to say in Section III about other topological single-particle problems such as the quasiparticle Hamiltonians of topological superconductors~\cite{schnyder,kitaev}.

Recent developments show that some important features of the 3D topological insulator can be understood in terms of the orbital contribution to magnetoelectric polarizability (Section IV), and more directly derived directly in 3D, or even by working down from 4D, than by working up from 2D; in this respect our discussion reflects later improvements in theoretical understanding.  
Some rapidly developing areas are reviewed in Section V, such as proposed proximity effects between superconductors and topological insulators.

\section{BASIC PHENOMENA AND EXPERIMENTS}

This section summarizes the current body of experimental knowledge on 3D topological insulators. Readers seeking to understand the theoretical background of the experimental observations may wish to consult Section III in parallel.  The first 3D topological insulator to be experimentally discovered was the semiconducting alloy Bi$_{1-x}$Sb$_{x}$, whose unusual surface bands were mapped in an angle-resolved photoemission (ARPES) experiment.  ARPES uses a photon to eject an electron from a crystal, then determines the surface or bulk electronic structure from an analysis of the momentum of the emitted electron.  The 3D topological state in Bi$_{1-x}$Sb$_{x}$ is a new state in the sense that it cannot be reduced to a quantum spin Hall state~\cite{moore&balents-2006,fu&kane&mele-2007,rroy3D}. Its surface is a new form of 2D metal where electron's spin and linear momentum are locked one to one and the two dimensional Fermi surface carries a nontrivial Berry's phase of $\pi$~\cite{fu&kane&mele-2007}. Its uniqueness can be understood in the following way : a layered three dimensional version of the quantum spin Hall insulator Hg(Cd)Te would be a "weak 3D topological insulator (WTI)" with invariant structure of ($\nu$$_o$=0,111). A WTI would exhibit surface states that contain even number of Dirac cones thus a net Berry's phase of 0 or 2$\pi$. The surface states of a WTI are not robust against disorder.

Unlike in a transport experiment, ARPES carried out in a spin resolution mode can, in addition, measures the distribution of spin orientations on the Fermi surface which can be used to estimate the Berry's phase on the surface. Spin sensitivity is critically important for probing the existence spin-momentum locking on the surface expected as a consequence of bulk topological order. The 3D topological insulator Bi$_{1-x}$Sb$_{x}$ on the other hand have odd number of surface states whose spin-texture supports a non-trivial Berry's phase providing protection against non-magnetic disorder. Such a topological insulator is described by ($\nu$$_o$=1, $\vec{M}$).

\begin{figure}
\includegraphics[scale=0.55]{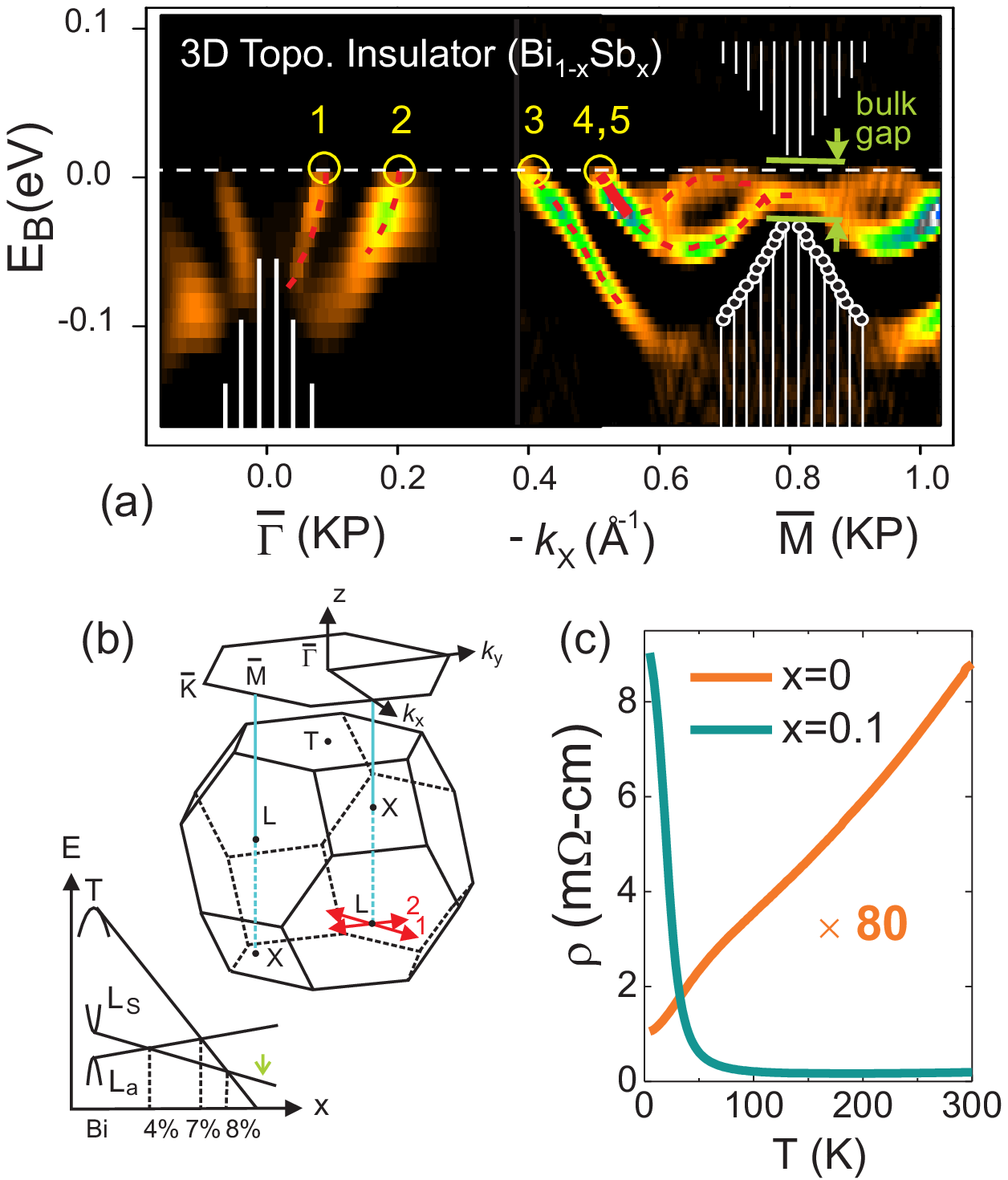}
\caption{\label{fig:Fig2b} \textbf{Topological surface states in Bi$_{1-x}$Sb$_{x}$:} Topological insulator Bi-Sb alloy exhibits odd number of surface state crossings between any pair of Kramers' points. (a) Bi-Sb alloy in the semiconducting composition range (0.07 $<$ x $<$ 0.18) exhibits odd number of surface state Fermi crossings in between any pair of Kramers' points. Bi$_{0.9}$Sb$_{0.1}$ features 5 such Fermi crossings where each band is non-degenerate. Surface state crossings along one such Kramers' pair $\Gamma$-M are shown. Detailed ARPES results suggest that semiconducting Bi-Sb alloys belong to the Z$_2$ = -1 topological class~\cite{hsieh} and it is the only known topological insulator that has non-trivial weak invariants~\cite{teo09}. (b) A schematic of the 3D Brillioun zone and its (111) hexagonal surface projection is shown. (c) Resistivity profile of pure Bi (semimetallic) is compared with semiconducting Bi-Sb alloy.}
\end{figure}

Bismuth antimony alloys have long been studied for their thermoelectric properties~\cite{lenoir}. Bismuth is a semimetal with strong spin-orbit interactions. Its band structure features an indirect negative gap between the valence band maximum at the T point of the bulk Brillouin zone (BZ) and the conduction band minima at three equivalent L points~\cite{liuallen}. The valence and conduction bands at L are derived from antisymmetric (L$_a$) and symmetric (L$_s$) $p$-type orbitals, respectively, and the effective low-energy Hamiltonian at this point is described by the (3+1)-dimensional relativistic Dirac equation~\cite{wolff}. The resulting dispersion relation, $E(\vec{k}) = \pm \sqrt{ {(\vec{v} \cdot \vec{k})}^2 + \Delta^2} \approx \vec{v} \cdot \vec{k}$, is highly linear owing to the combination of an unusually large band velocity $\vec{v}$ and a small gap $\Delta$ (such that $\lvert \Delta / \lvert \vec{v} \rvert \rvert \approx 5 \times 10^{-3} $\AA$^{-1}$) and has been used to explain various peculiar properties of bismuth. Substituting bismuth with antimony changes the critical energies of the band structure.  At an Sb concentration of $x \approx 4\%$, the gap $\Delta$ between L$_a$ and L$_s$ closes and a massless three-dimensional (3D) Dirac point is realized. As $x$ is further increased this gap re-opens with inverted symmetry ordering, which leads to a change in sign of $\Delta$ at each of the three equivalent L points in the BZ. For concentrations greater than $x > 7\%$ there is no overlap between the valence band at T and the conduction band at L, and the material becomes an inverted-band insulator. Once the band at T drops below the valence band at L, at $x \approx 7\%$, the system evolves into a direct-gap insulator with a massive Dirac-like bulk bands.

\begin{figure}
\includegraphics[scale=0.4]{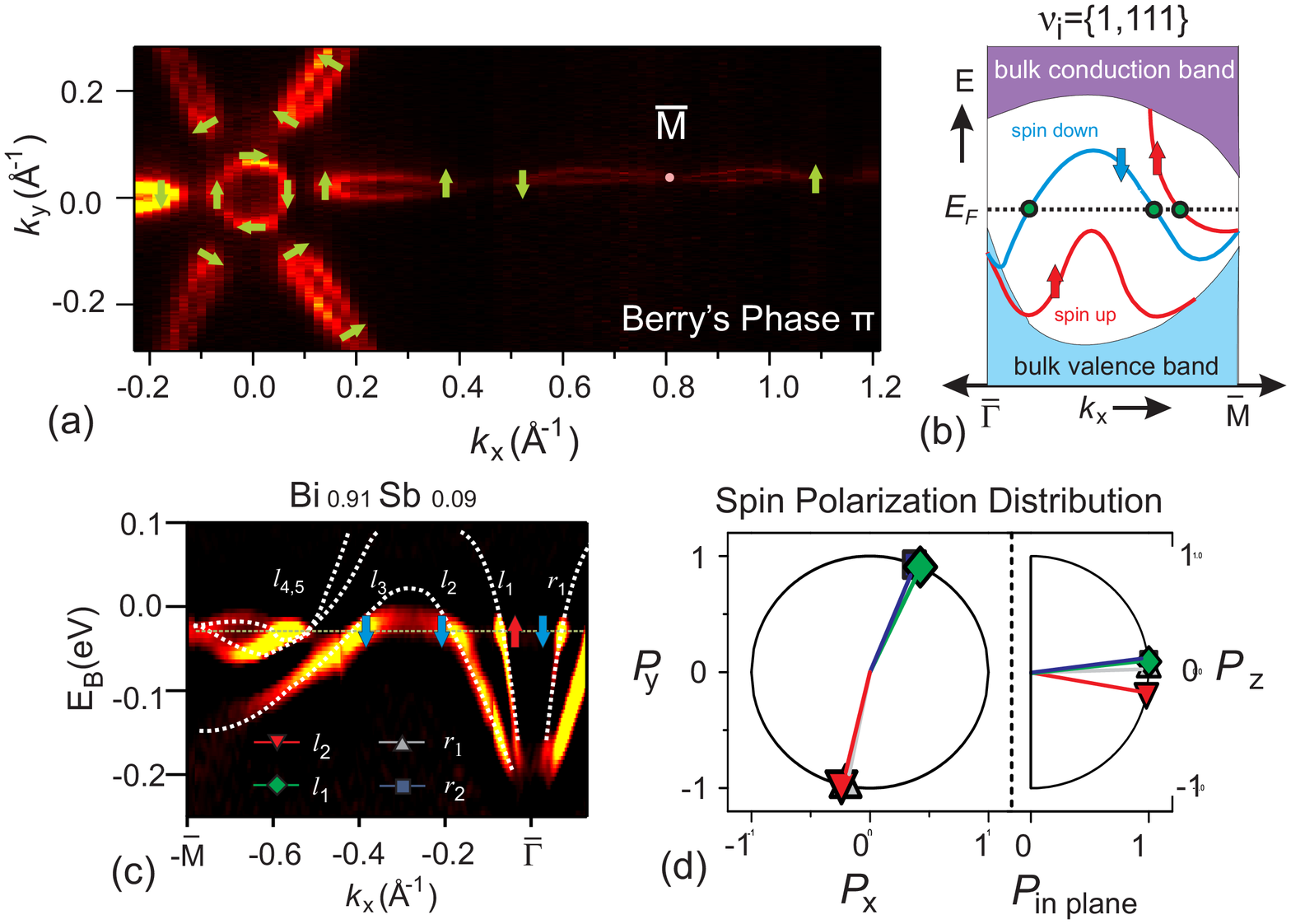}
\caption{\label{fig:Fig2} \textbf{Fermi surface spin-textures:} Topological surfaces exhibit non-trivial spin-textures. Spin-resolved photoemission directly probes the topological character of the spin-textures of the surface. (a) A schematic of spin-ARPES measurement set up is shown which was used to measure the spin distribution mapped on the (111)-surface Fermi surface of Bi$_{0.91}$Sb$_{0.09}$. Spin orientations on the surface create a vortex-like structure around $\Gamma$-point. A net Berry's phase $\pi$ is extracted from the full Fermi surface data measured over the Bi$_{1-x}$Sb$_{x}$ to Sb. (b) A minimal topology of the band-structure is shown in an schematic based on the systematic data. (c) Relative spin distributions are shown per surface band. (d) Net polarization along x-, y- and z- directions are shown. \textit{P$_z$}$\sim$0 suggests that spins lie mostly within the surface plane (Adapted from Hsieh \emph{et.al, }~\cite{hsieh09a}.}
\end{figure}

Although the presence of bulk and surface states in Bi-based materials were known, experiments by Hsieh \emph{et.al.}~\cite{hsieh} establish the existence of a direct energy gap in the bulk Dirac-like band-structure of Bi$_{0.9}$Sb$_{0.1}$. The linearly dispersive bulk bands are uniquely consistent with strong spin-orbit coupling nature of the inverted insulator. The same experiments observed several non-degenerate surface states that span the bulk gap. In general, the states at the surface of spin-orbit coupled compounds are allowed to be spin split owing to the loss of space inversion symmetry $[E(k,\uparrow) = E(-k,\uparrow)]$. However, as required by Kramers' theorem, this splitting must go to zero at the four time reversal invariant momenta (TRIM) in the 2D surface BZ as required by Kramers theorem. Along a path connecting two TRIM in the same BZ, the Fermi energy inside the bulk gap will intersect these singly degenerate surface states either an even or odd number of times. When there are an even number of surface state crossings, the surface states are topologically trivial because weak disorder (as may arise through alloying) or correlations can remove \emph{pairs} of such crossings by pushing the surface bands entirely above or below $E_F$. When there are an odd number of crossings, however, at least one surface state must remain gapless, which makes it non-trivial.  Such ``topological metal'' surface states cannot be realized in a purely 2D electron gas system such as the one realized at the interface of GaAs/GaAlAs or on the surface of Si.

\begin{figure}
\includegraphics[scale=0.2]{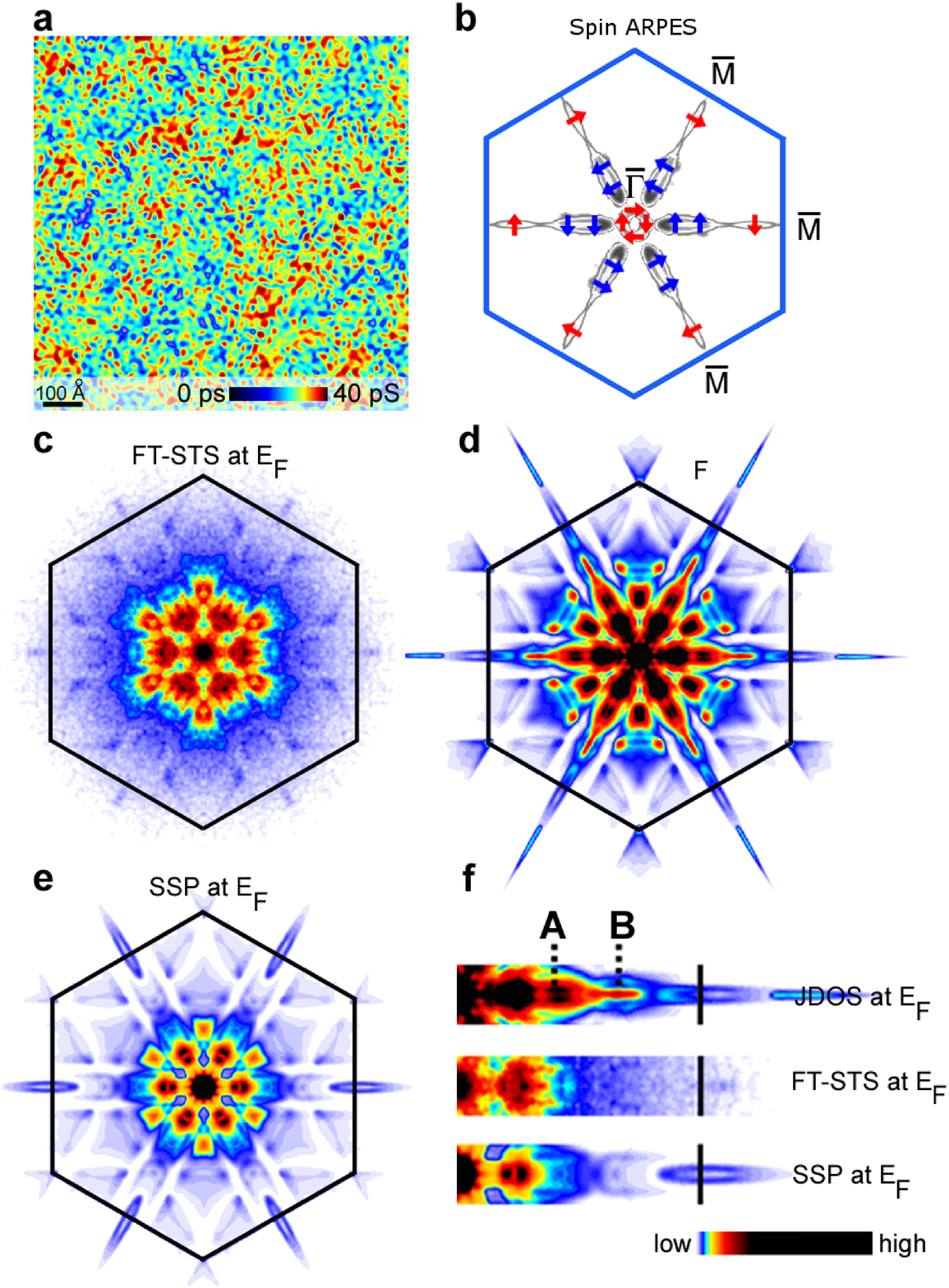}
\caption{\label{fig:FigSTM} \textbf{Absence of backscattering}: Quasiparticle Interference observed in Bi-Sb alloy surface exhibits absence of elastic backscattering: (a) Spatially resolved conductance maps of the Bi$_{0.92}$Sb$_{0.08}$ (111) surface obtained at 0 mV over a 1000$\AA$$\times$1000$\AA$. (b) ARPES intensity map of the surface state measured at the Fermi level. The spin textures from spin-ARPES measurements are shown with arrows, and high symmetry points are marked $\Gamma$ and three Ms). (c) The Fourier transform scanning tunneling spectroscopy (FT-STS) at E$_F$. (d) The joint density of states (JDOS) at E$_F$. (e) The spin-dependent scattering probability(SSP) at E$_F$. (f) Close-up view of the from JDOS, FT-STS and SSP at E$_F$, along the $\Gamma$-M direction. (Adapted from Roushan \emph{et.al.,}~\cite{roushan} and Hsieh \emph{et.al.,}~\cite{hsieh09a}.}
\end{figure}

On the (111) surface of Bi$_{0.9}$Sb$_{0.1}$, the four TRIM are located at $\bar{\Gamma}$ and three $\bar{M}$ points that are rotated by $60^{\circ}$ relative to one another. Owing to the three-fold crystal symmetry (A7 bulk structure) and the mirror symmetry of the surface Fermi surface across $k_x = 0$, these three $\bar{M}$ points are equivalent. In pure bismuth (Bi), two surface bands emerge from the bulk band continuum near $\bar{\Gamma}$ to form a central electron pocket and an adjacent hole lobe. These two bands result from the spin-splitting of a surface state and are thus singly degenerate. In Bi-Sb alloy, in addition, a narrow electron pocket which is doubly degenerate around $\bar{M}$ is also observed (Fig.1). This is indicated by its splitting below $E_F$ between $-k_x \approx 0.55 \mathring{A}^{-1}$ and $\bar{M}$, as well as the fact that this splitting goes to zero at $\bar{M}$ in accordance with Kramers theorem. In Bi$_{0.9}$Sb$_{0.1}$, the states near $\bar{M}$ fall completely inside the bulk energy gap preserving their purely surface character at $\bar{M}$ unlike in Bi. The surface Kramers point is located approximately 15 $\pm$ 5 meV below $E_F$ at $\vec{k}  = \bar{M}$. The observation of five surface state crossings (an odd rather than an even number) between $\bar{\Gamma}$ and $\bar{M}$, confirmed by the observation of the Kramers degenerate point confirms the topologically non-trivial nature of the material. While these results are consistent with a Z$_2$ topological insulator scenario for Bi-Sb, more decisive evidence comes from the study of spin-texture of the surface Fermi surface.

The bulk topological Z$_2$ order is expected to give rise to unusual spin physics at the edges or surfaces of topological insulators. However, unlike quantum Hall systems, the topological insulator surfaces do not necessarily exhibit a quantized charge or spin response. In fact, the spin polarization is not a conserved quantity in a spin-orbit material. Thus, their topological quantum numbers, the analogs of the Chern number, cannot be measured via the classic magneto-transport methods. ARPES with spin sensitivity can perform analogous measurements for topological insulators by mapping out all 4 topological quantum numbers that uniquely identify the topological class. Hsieh \emph{et.al.}~\cite{hsieh09a} measured the topological numbers for Bi$_{1-x}$Sb$_x$ and provided an identification of its spin texture, which heretofore was unmeasured despite its surface states (SSs) having been observed. The measured spin texture (Fig.2) reveals the existence of a nonzero geometrical quantum phase, a Berry's phase of $\pi$ and the handedness or chiral properties. Spin-ARPES technique enables to investigate aspects of the metallic regime of the Bi$_{1-x}$Sb$_x$ series, such as spin properties in pure Sb, which are necessary to determine the microscopic origin of topological order. Spin-ARPES measurements on Sb showed that its surface carries a geometrical (Berry's) phase and chirality property unlike the conventional spin-orbit metals such as gold, which has zero net Berry's phase and no net chirality.

\begin{figure}
\includegraphics[scale=0.5]{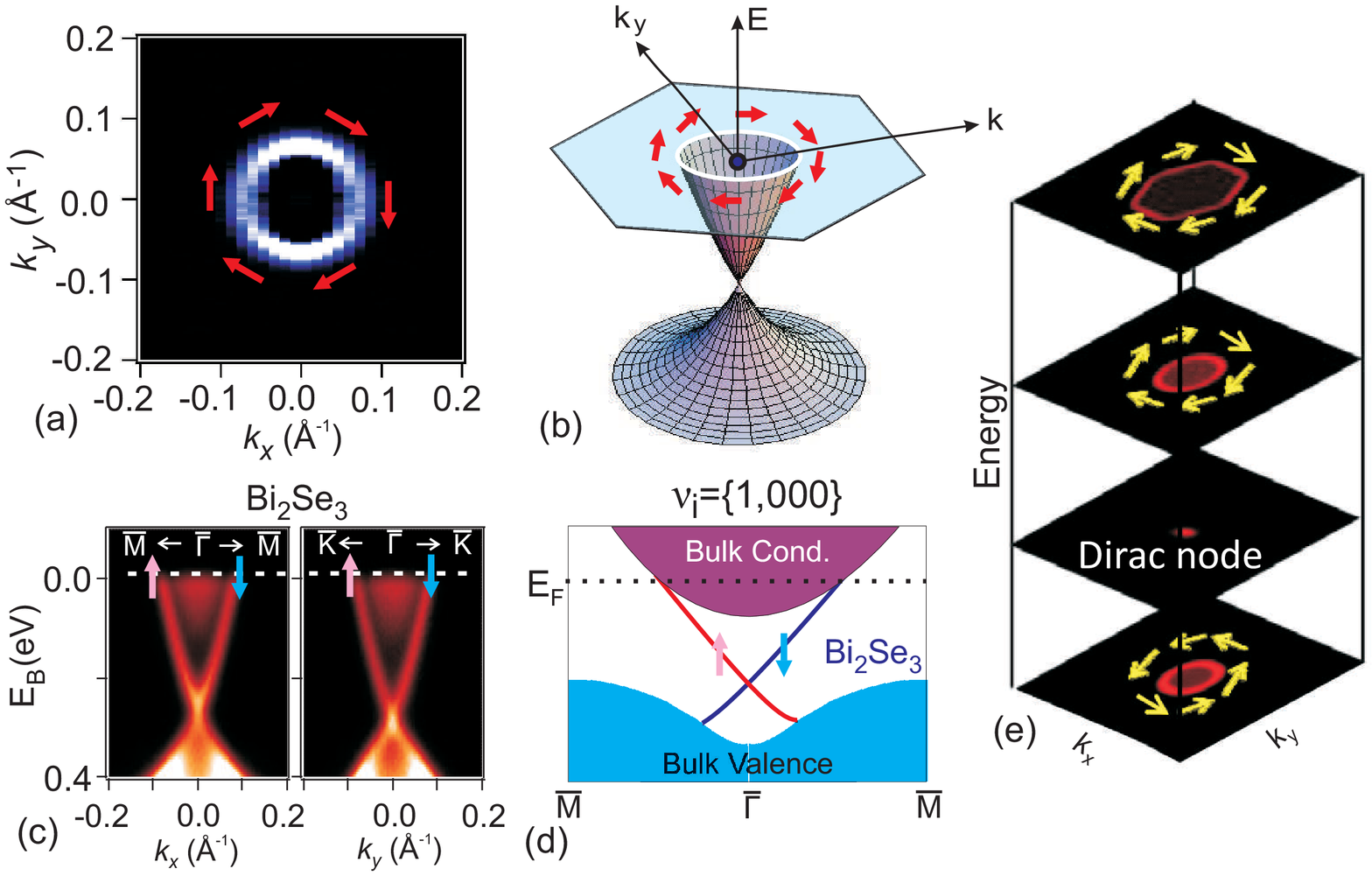}
\caption{\label{fig:Fig5v2} \textbf{Single Dirac cone fermions}: Spin-momentum locked helical surface Dirac fermions are hallmark signatures of bulk topological Z$_2$ order. The single-Dirac-cone per surface topological insulators belong to the ($\nu$$_o$=1,000) topological class. (a) Surface dispersion relation of electronic states reveal a spin-polarized Dirac cone. (b) Surface Fermi surface exhibits chiral left-handed spin textures. Schematic presentations of (c) surface and bulk band topology and (d) a single spin-polarized Dirac fermion on the surface of Bi$_2$X$_3$ ($\nu$$_o$=1,000) topological insulators. (e) Constant-energy deposition maps of the surface state are labeled with spin texture.  (Adapted from Xia \emph{et.al.,}~\cite{xia09a,xiaarxiv}; Hsieh \emph{et.al.,}~\cite{hsieh09b})}
\end{figure}


The odd number of Fermi crossings protects the topological surface from becoming insulating, regardless of the position of the chemical potential or the influence of non-magnetic perturbations. Spin-ARPES experiments have established that these surface states have a chiral spin structure and carries a non-trivial Berry's phase, which makes
them distinct from ordinary surface states. All these characteristics suggest that backscattering, or scattering between states of equal and opposite momentum, which results in Anderson localization in typical low-dimensional systems~\cite{anderson}, will not occur for these two-dimensional carriers. Random alloying in Bi$_{1-x}$Sb$_x$, which is not present in other material families of topological insulators found to date, makes this material system an ideal candidate in which to examine the impact of disorder or random potential on topological surface states. The fact that the chiral two-dimensional states are indeed protected from spin-independent scattering was established by Roushan \emph{et.al.}~\cite{roushan} by combining results from scanning tunneling spectroscopy (STM/STS) and spin-ARPES on Bi$_{0.92}$Sb$_{0.08}$. The results show that despite strong atomic scale disorder and random potential, elastic-backscattering between states of opposite momentum and opposite spin is absent. These observations demonstrate that the chiral nature of these states protects the spin of the carriers which is ultimately a consequence of time-reversal symmetry. The same conclusion has emerged from studies of the electronic interference patterns near defects or steps on the surface that electrons never turn completely around when scattered in other topological insulators~\cite{tzhang,alpichshev}).

While the surface structure of Bi$_{1-x}$Sb$_x$ alloy material was found to be rather complicated and bandgap to be rather small, this work launched a search for topological insulators with larger bandgap and simpler surface spectrum for observing topological phenomena at high temperatures. Topological phenomena in materials are usually very fragile and impossible to create without liquid helium temperatures and high magnetic fields. New three-dimensional "topological insulator" materials, especially Bi$_2$Se$_3$, offer the potential for protected surface states and other topological behavior at room temperature. In 2008, work led by a Princeton group used angle-resolved photoemission spectroscopy (ARPES) and first-principles calculations to study the surface band structure of Bi$_2$Se$_3$ and observe the characteristic signature of a topological insulator in the form of a single-Dirac-cone~\cite{xia09a,xiaarxiv}. Concurrent theoretical work by H.Zhang \emph{et al.}~\cite{zhangn} used electronic structure methods to show that Bi$_2$Se$_3$ is just one of several new large-bandgap topological insulators. Zhang \emph{et.al.,} also provided a simple tight-binding model to capture the single-Dirac-cone observed in Bi$_2$Te$_3$~\cite{noh} and made new predictions about Bi$_2$Se$_3$ and Sb$_2$Te$_3$. Subsequently, detailed and systematic surface investigations of Bi$_2$Se$_3$~\cite{hsieh09b,hor09,park10}; Bi$_2$Te$_3$~\cite{chen,hsieh09b,hsieh09c,xia09b} and  Sb$_2$Te$_3$~\cite{hsieh09c} confirmed the topological band-structure of all 3 of these materials.

\begin{figure}[t]
\includegraphics[scale=0.18]{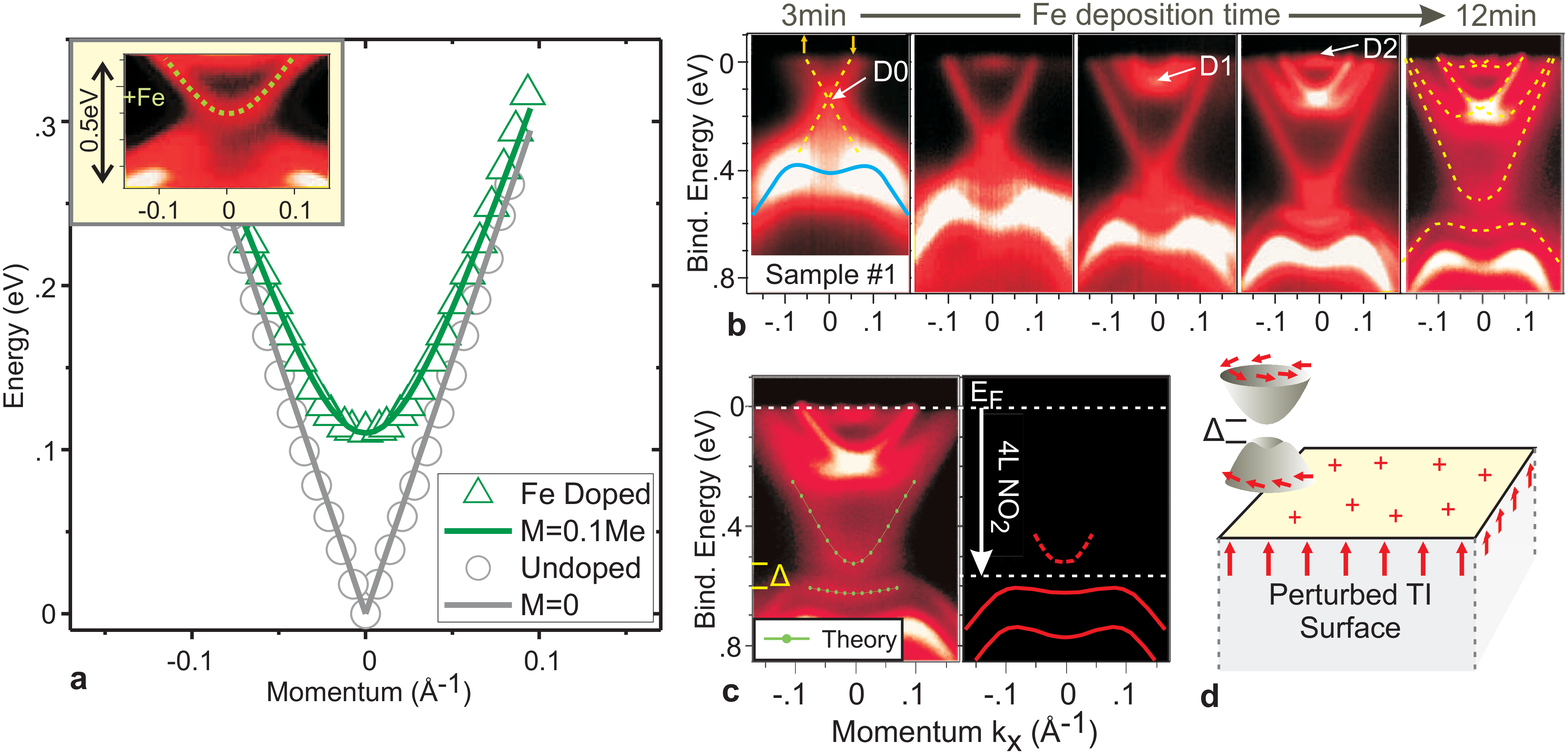}
\caption{{\bf{Surface Dirac points are protected by time reversal symmetry}}:
(a) The surface state dispersion is compared before and after deposition, using hollow circles
to indicate data and lines to trace ideal Dirac dispersions. After deposition, the upper Dirac cone gains a mass of 0.1 electron masses. (b) The evolution of Bi2Se3 surface states as a function of surface Fe deposition is shown for a p-type Bi$_2$Se$_3$ sample hole doped with 0.25\% Ca. (c, left) A z-axis magnetic perturbation causes (green) GGA predicted surface states to conform to the iron doped dispersion. (right) A surface insulator phase can likely be obtained by adding a non-magnetic p-type surface dopant such as NO$_2$ subsequent to Fe deposition. (d) A cartoon illustrates the charge and magnetic perturbations caused by Fe deposition on the TI surface.}
\end{figure}


The unusual planar metal that forms at the surface of topological insulators inherits topological properties from the bulk bandstructure. The manifestation of this bulk-surface connection occurs at a smooth surface where momentum along the surface remains well-defined: in its simplest possible form, each momentum along the surface has only a single spin state at the Fermi level, and the spin direction rotates as the momentum moves around the Fermi surface ensuring a non-trivial Berry's phase. These two defining properties of topological insulators namely spin-momentum locking of surface states and $\pi$ Berry's phase along with the consequences such as the robustness to non-magnetic disorder could be most clearly demonstrated with the discovery of the second generation of topological insulators.

\begin{figure}
\includegraphics[scale=0.31]{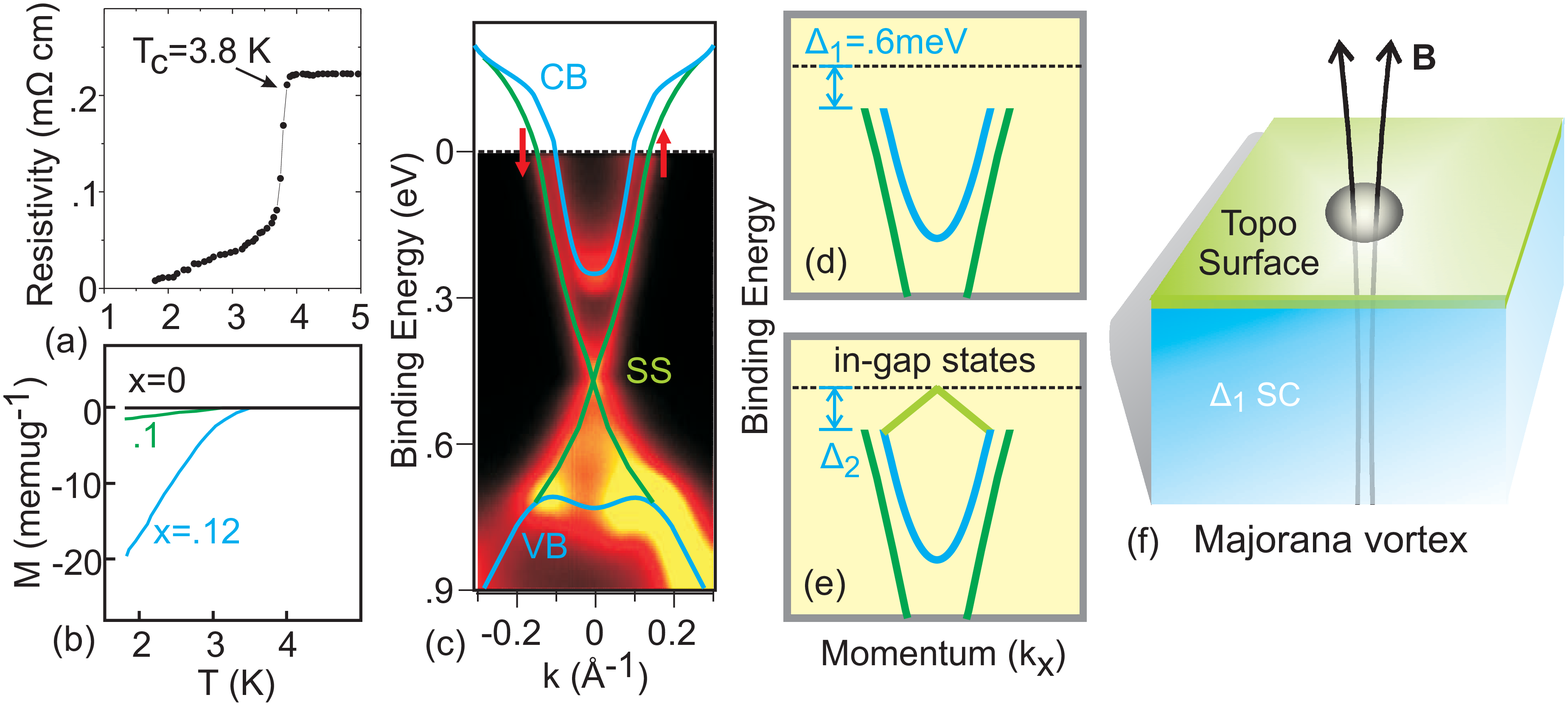}
\caption{\label{fig:Fig5v3} \textbf{Superconductivity in doped topological insulators}: (a) Superconductivity in Cu-doped Bi2Se3 with transition temperature of 3.8K. (b) Meissner transition. (c) Topologically protected surface states cross the Fermi level before merging with the bulk valence and conduction bands in superconducting Cu$_{0.12}$Bi$_2$Se$_3$. (d) If the superconducting wavefunction has even parity, the Fermi level will be fully gapped and vortices on the crystal surface will host braidable Majorana fermions. (e) If superconducting parity is odd ($\Delta_2$ state), the material will be a ``topological superconductor", and new gapless states will appear below T$_C$. (f) Majorana fermion surface vortices are found at the end of bulk vortex lines and might be adiabatically manipulated for quantum computation if superconducting pairing is even ($\Delta_1$)}
\end{figure}

While the phase observed in Bi$_{1-x}$Sb$_x$ alloys is theoretically the same as that observed in the Bi$_2$Se$_3$ class, there are three crucial differences that suggest that this series may become the reference material for future experiments on this phase. The Bi$_2$Se$_3$ surface state is found from ARPES and theory to be nearly the idealized single Dirac cone as seen from the experimental data in Fig.4\&5. Second, Bi$_2$Se$_3$ is stoichiometric (i.e., a pure compound rather than an alloy like Bi$_{1-x}$Sb$_x$) and hence can be prepared in principle at higher purity.  While the topological insulator phase is predicted to be quite robust to disorder, many experimental probes of the phase, including ARPES of the surface band structure, are clearer in high-purity samples. Finally, and perhaps most important for applications, Bi$_2$Se$_3$ has a large band gap of approximately 0.3 eV (equivalent to 3600 Kelvin). In its high purity form, the large band gap of Bi$_2$Se$_3$ indicates that topological behavior can be seen at room temperature and greatly increases the potential for applications (Fig.7). To understand the likely impact of these new topological insulators, an analogy can be drawn with the early days of high-temperature cuprate superconductivity: the original cuprate superconductor LBCO was quickly superseded by "second-generation" materials such as YBCO and BSCCO for most applied and scientific purposes~\cite{joelrev}.

\begin{figure}
\includegraphics[scale=0.5]{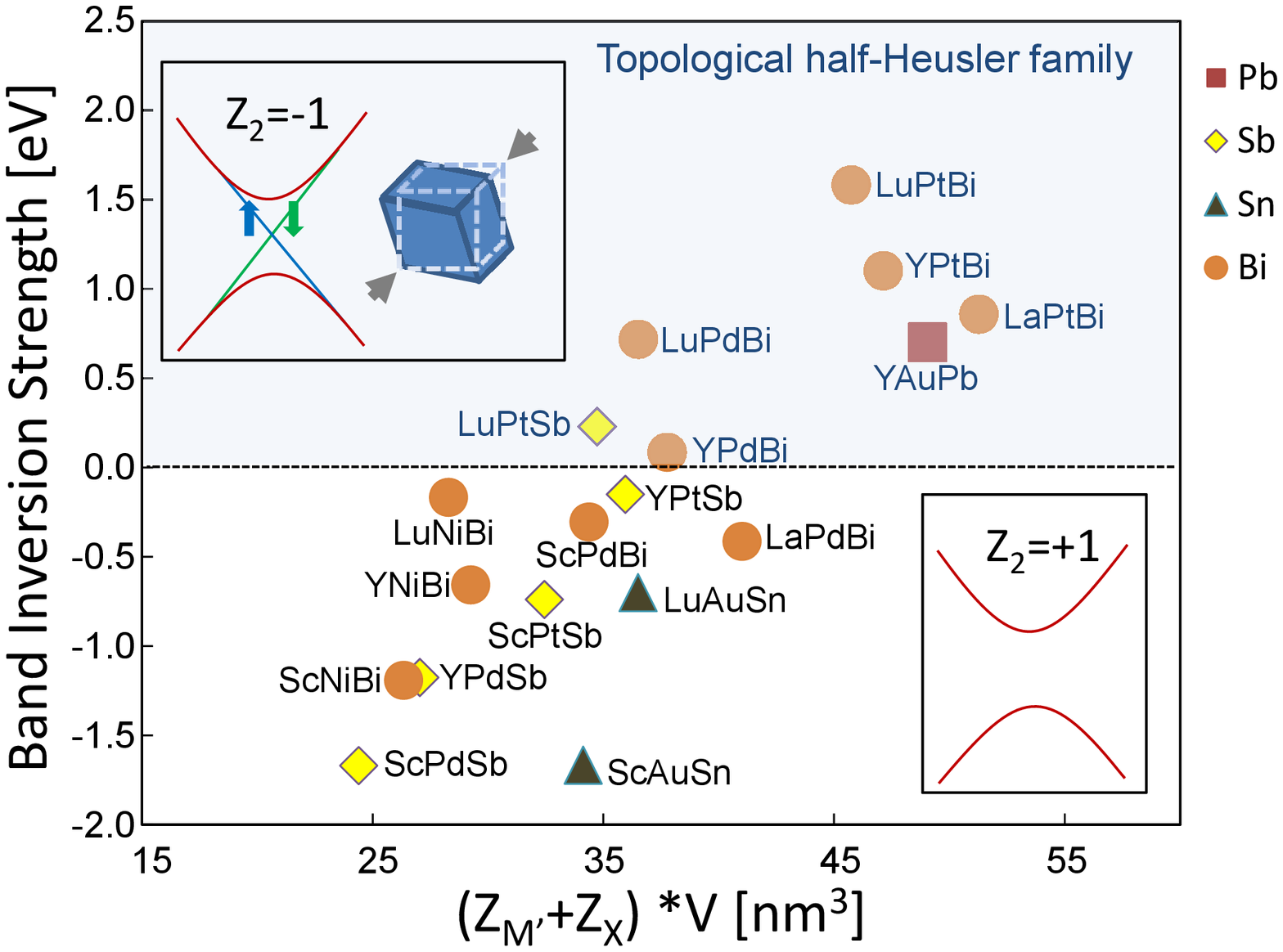}
\caption{\label{fig:Fig7exp} \textbf{Half-Heusler family of topological insulators} Topological band inversion strengths are plotted as a function of the product of the unit-cell volume (V)and the sum of the nuclear charges of M' and X atoms for several classes of half-Heuslers. The inversion strength is defined as the energy difference between the G8 and G6 states at the G point. Negative values denote absence of inversion. Materials with positive band inversion strength are candidates for topological insulators with Z$_2$=-1 once the lattice symmetry is broken. LuPtBi, YPtBi and LaPtBi possess the most stable inverted Z$_2$=-1 type bulk band structures. (Lin \emph{et.al.,}~\cite{lin2010}) }
\end{figure}

All the key properties of topological states have been demonstrated for Bi$_2$Se$_3$ which has the most simple Dirac cone surface spectrum and the largest band-gap (0.3 eV). In Bi$_2$Te$_3$ the surface states exhibit large deviations from a simple Dirac cone due to a combination of smaller band-gap (0.15 eV) and a strong trigonal potential~\cite{chen} which can be utilized to explore some aspects of its surface properties~\cite{fuwarping,hasan09}. Hexagonal deformation of the surface states is confirmed by STM measurements~\cite{alpichshev}. Speaking of applications within this class of materials, Bi$_2$Te$_3$, is already well known to materials scientists working on thermoelectricity: it is a commonly used thermoelectric material in the crucial engineering regime near room temperature.



Two defining properties of topological insulators namely spin-momentum locking of surface states and $\pi$ Berry's phase could be clearly demonstrated in Bi$_2$X$_3$ series. The protected nature of the surface states are also most straightforwardly demonstrated. Topological surface states are protected by time reversal symmetry which implies that the surface state Dirac node would be robust in the presence of non-magnetic disorder but open a gap in the presence of time-reversal breaking perturbation of the system such as the effect of magnetic impurities and the associated disorder. Magnetic impurity such as Fe on the surface of Bi$_2$Se$_3$ opens a gap at the Dirac point (Fig.5)~\cite{xiaarxiv,hsieh09b,wray10}.  The magnitude of the gap is likely set by the interaction of Fe ions with the Se surface and the time-reversal breaking disorder potential introduced on the surface. Non-magnetic disorder created via molecular absorbent NO$_2$ or alkali atom adsorption (K or Na) on the surface leaves the Dirac node intact (Fig.5) in both Bi$_2$Se$_3$ and Bi$_2$Te$_3$~\cite{xia09b,hsieh09b}. These results are consistent with the fact that the topological surface states are protected to be metallic by time-reversal symmetry.

Most of the interesting theoretical proposals that utilize topological insulator surfaces require the chemical potential to lie at the degenerate Dirac node similar to the case in graphene. In graphene, the chemistry of carbon atoms naturally locates the Fermi level at the Dirac point --- the point at which the two cones intersect --- where the density of states vanishes. This makes its density of carriers highly tunable by an applied electrical field and enables applications of graphene to both basic science and microelectronics. The surface Fermi level of a topological insulator does not have any particular reason to sit at the Dirac point, but by a combination of surface and bulk chemical modification, tuning to the Dirac point in Bi$_2$Se$_3$ was demonstrated by Hsieh et.al.~\cite{hsieh09b}. This was achieved by passivating the Se vacancies with Ca in the bulk to place the Fermi level within the bulk bandgap and then hole doping the surface with NO$_2$ to keep the chemical potential at the Dirac point. While this is an important advancement for furthering the spectroscopic measurements, it needs to be seen how such a degree of surface engineering can be achieved in a transport measurement geometry.

Recent observations of superconductivity in Cu$_x$Bi$_2$Se$_3$ (T$_c$ = 3.8K) by Hor \emph{et.al.,} ~\cite{cavasc} have raised interest in understanding the role of spontaneously broken symmetry within a topological band-structure. Measurements of electronic band structure in Cu$_x$Bi$_2$Se$_3$ (x=0 to 12\%) by Wray \emph{et.al.,}~\cite{wray09} suggest that superconductivity occurs with electrons in a bulk relativistic kinematic regime and we identify that an unconventional doping mechanism causes the nontrivial topological surface character of the undoped compound to be preserved at the Fermi level of the superconducting compound, where Cooper pairing occurs. Moreover, they report key band parameters that constrain the form of the superconducting state. These experimental observations provide important clues for developing a theory of superconductivity in doped Bi-based topological insulators. Two possible theoretical scenarios are shown in Fig.6.

While the 3D topological insulators, whose variants can host magnetism and superconductivity, has generated widespread research activity in condensed-matter and materials-physics communities, there has also been intense efforts to discover new materials class. This is largely because some of the most interesting topological phenomena, however, require topological insulators to be placed in multiply connected, highly constrained geometries with magnets and superconductors all of which thus require a large number of functional variants with materials design flexibility as well as electronic, magnetic and superconducting tunability. Given the optimum materials, topological properties open up new vistas in spintronics, quantum computing and fundamental physics. We have extended the search for topological insulators from the binary Bi-based series to the ternary thermoelectric Heusler compounds. It has been recently found that although a large majority of the well-known Heuslers such as TiNiSn and LuNiBi are rather topologically trivial, the distorted LnPtSb-type (such as LnPtBi or LnPdBi, Ln=fn lanthanides) compounds belonging to the half-Heusler subclass harbour Z$_2$=-1 topological insulator parent states, where Z$_2$ is the band parity product index. Half-Heuslers provide a new platform for deriving a host of topologically exotic compounds and their nanoscale or thin-film device versions through the inherent flexibility of their lattice parameter, spin–orbit strength and magnetic moment tunability paving the way for the realization of multifunctional topological devices~\cite{lin2010,chadov10}.  Heuslers are not insulating in their native state but growing them on substrates making thinfilms or quantum-well structures provide a way to open a band-gap turning them into real insulators.

While in the above we have focused on crystals for basic characterization, films and nanostructures of topological insulators will eventually be important for applications and have already enabled important tests of theoretical predictions.  Two examples are the observation of a thickness-dependent gap in MBE-grown films~\cite{gapexp} using in-situ ARPES and the transport measurement of Aharonov-Bohm oscillations in topological insulator nanoribbons~\cite{cui} (for the latter, see also the theoretical works~\cite{bardarson,zhangvishwanath}).  In addition to applications to thermoelectrics and spintronics~\cite{gapp}, such thin films might enable observation of an unusual exciton condensate~\cite{franzmoore}.  The main remaining limitation of current materials, especially for experimental techniques that (unlike ARPES) do not distinguish directly between bulk and surface states, is that they have some residual conduction in the bulk from impurity states even if the band-structure features an energy gap.




\section{INDEPENDENT-ELECTRON THEORY OF 3D TOPOLOGICAL INSULATORS}

One view of topological insulators is that they represent a continuation, and possibly the culmination, of a long-term program, initiated by David Thouless and collaborators~\cite{tknn}, to understand the consequences of the Berry phase for electrons in solids.  Their first result, even before the work by Berry that explained the general concept of the Berry phase~\cite{berry}, related the topology of electron wavefunctions in the to the integer quantum Hall effect.  Topological phases in which a response function is expressed in terms of a topological invariant are often referred to as ``Thouless-type topological phases''; the IQHE and topological insulators in 2D and 3D are examples.

The integer quantum Hall effect has the remarkable property that, even at finite temperature in a disordered material, a transport quantity is quantized to remarkable precision: the transverse (a.k.a. Hall) conductivity is $\sigma_{xy} = n e^2 / h$, where $n$ is integral to 1 part in $10^9$.  This quantization results because $\sigma_{xy}$ is determined by a topological invariant, and understand this intrinsically 2D topological invariant also leads to the modern theory of polarization, which is mathematically a 1D quantity (i.e., it exists in dimensions 1 and above).  We review these developments in order to understand the single-particle picture of 3D topological insulators and the connection to orbital magnetoelectric polarizability.  Our discussion starts with noninteracting electrons in a perfect crystal and then explain the generalizations to disorder and interactions.


In the IQHE and topological insulators, the topological invariant can be obtained from integration over the Brillouin zone of a quantity derived from the Berry phase of electron wavefunctions.  The idea of the Berry phase~\cite{berry} is that, as an eigenstate evolves under a Hamiltonian whose parameters change adiabatically, there is a physically significant phase that accumulates.  The Berry phase can be rather important even when it is not part of a topological invariant.  In crystalline solids, the electrical polarization, the anomalous Hall effect, and one part of the magnetoelectric polarizability all derive from ``non-topological'' Berry phases of the Bloch electron states.

 
 The space of parameters for these Berry phases is just the crystal momentum ${\bf k}$: Bloch's theorem for electrons in a periodic potential states that, given a potential invariant under a set of lattice vectors ${\bf R}$, $V({\bf r} + {\bf R}) = V({\bf r})$, the electronic eigenstates can be labeled by a ``crystal momentum'' ${\bf k}$ and written in the form
\beq
\psi_{\bf k}({\bf r}) = e^{i {\bf k} \cdot {\bf r}} u_{\bf k} ({\bf r}),
\eeq
where the function $u$ has the periodicity of the lattice.  Note that the crystal momentum ${\bf k}$ is only defined up to addition of reciprocal lattice vectors, i.e., vectors whose dot product with any of the original lattice vectors is a multiple of $2 \pi$.

While the energetics of Bloch wavefunctions (``band structure'') underlies many properties of solids, and matrix elements of the wavefunctions are important for optical and other properties, it has been appreciated only recently that many phenomena are governed by Berry-phase physics arising from the smooth dependence of the wavefunction $u_k$ on $k$.  Even in a one-dimensional system, there is a nontrivial ``closed loop'' in the parameter $k$ that can be defined because of the periodicity of the ``Brillouin zone''' $k \in [-\pi/a,\pi/a)$:
\beq
\gamma = \oint_{-\pi/a}^{\pi/a}\,\langle u_k | i \partial_k | u_k \rangle dk.
\eeq
Like other Berry phases, the phase $\gamma$ is obtained as an integral of a ``connection'' ${\cal \bf A} = \langle u_k | i \partial_k | u_k \rangle$.  In this 1D example, ${\cal \bf A}$ is just a scalar; in higher dimensions it is a vector potential that lives in momentum space.
How are we to interpret this Berry phase
physically?  An intuitive clue is provided if we make the replacement $i \partial_k$ by $x$, as would be appropriate if we consider the action on a plane wave.  This suggests, correctly, that the Berry phase may have something to do with the spatial location of the electrons: in fact the electronic polarization is just $e \gamma / 2 \pi$~\cite{ksv}.

Another clue to what the phase $\gamma$ might mean physically is provided by asking if it is gauge-invariant.  In the usual situation, gauge-invariance of the Berry phase results from assuming that the wavefunction could be continuously defined on the interior of the closed path.  But here we have a closed path on a noncontractible manifold; the path in the integral winds around the Brillouin zone, which has the topology of the circle.  (Here we are implicitly introducing another kind of topological invariant.  One kind comes from integrating differential forms derived from the Berry phase, which is a ``cohomology'' invariant; another kind are ``homotopy'' invariants, the generalizations of the notion of winding number.  The homotopy groups $\pi_n(M)$, which describe equivalence classes of mappings from the sphere $S^n$ to the manifold $M$, will occasionally be mentioned in technical remarks below.  The book of Nakahara~\cite{nakahara} is a physics-oriented reference that covers both cohomology and homotopy.)

What happens to the Berry phase if we introduce a phase change $\phi(k)$ in the wavefunctions, $|u_k\rangle \rightarrow e^{-i \phi(k)} |u_k\rangle$, with $\phi(\pi/a) = \phi(-\pi/a)+2 \pi n, n \in {\mathbb Z}$?  Under this transformation, the integral shifts as
\beq
\gamma \rightarrow \gamma + \oint_{-\pi/a}^{\pi/a} (\partial_k \phi) \, dk = \gamma + 2 \pi n.
\eeq
So redefinition of the wavefunctions shifts the Berry phase, and this corresponds to changing the polarization by a multiple of the ``polarization quantum'', which in one dimension is just the electron charge.  (In higher dimensions, the polarization quantum of a crystal is one electron charge per transverse unit cell.)  Physically the ambiguity of polarization corresponds to how adding an integer number of charges to one allowed termination gives another allowed termination~\cite{resta}.  The Berry phase is not gauge-invariant, but any fractional part it had in units of $a$ {\it is} gauge-invariant.
We discussed this one-dimensional example in some detail because topological insulators will be very similar.  The above calculation suggests that, to obtain a fully gauge-invariant quantity, we need to consider a two-dimensional crystal rather than a one-dimensional one.  Then integrating the Berry curvature, rather than the Berry connection, has to give a well-defined gauge-invariant quantity.

Indeed, the one-dimensional polarization example is related to the two-dimensional integer quantum Hall effect, which historically came first.  When a lattice is put in a commensurate magnetic field (one with rational flux per unit cell, in units of the flux quantum so that Bloch's theorem applies), each occupied band $j$ contributes an integer~\cite{tknn}
\beq
n_j = {i \over 2 \pi} \int\,dk_x\,dk_y\,\left(\langle \partial_{k_x} u_j | \partial_{k_y} u_j \rangle - \langle \partial_{k_y} u_j | \partial_{k_x} u_j \rangle \right)
\label{tknn}
\eeq
to the total Hall conductance:
\beq
\sigma_{xy} = {e^2 \over h} \sum_{j} n_j.
\eeq
This topological quantity (the ``TKNN integer'' or ``Chern number'', expressed as an integral over the Berry flux, which is the curl of the Berry connection $A^j = i \langle u_j | \nabla_k u_j \rangle$) is connected to the IQHE and also changes in polarization.  The relationship between polarization in 1D, which has an integer ambiguity, and the IQHE in 2D, which has an integer quantization, is the simplest example of the relationship between Chern-Simons forms in odd dimension and Chern forms in even dimension.  Looking ahead, the topological insulator in 3D can be similarly related to the topological invariants in 4 dimensions~\cite{asss} that underlie the ``four-dimensional quantum Hall effect''~\cite{zhanghu}.

One might worry whether the TKNN integer defined in equation (\ref{tknn}) is specific to noninteracting electrons in perfect crystals.  An elegant way to generalize the definition physically, while keeping the same mathematical structure, was developed by Niu, Thouless, and Wu~\cite{niuthouless}.  One can imagine a disordered system on a torus and rewrite the topological invariant above in terms of how much charge is ``pumped'' around one circle by adiabatic insertion of flux through the other circle.  The same trick can be used to generalize the 3D topological insulator invariants described below to disordered systems.  While for the IQHE the ``flux trick'' is sufficient to handle interacting systems as well as disordered ones, a different approach is necesary for topological insulators.  One final note before moving on to time-reversal-invariant Fermi systems is that, while we have concentrated above on giving explicit expressions for the invariants in terms of wavefunctions, which is useful to get their physical meaning, it is often more direct for topological classification to consider homotopy groups of the Bloch Hamiltonians~\cite{ass,asss}.

\subsection{Time-reversal invariance in Fermi systems}

Kane and Mele showed in 2005 that imposing time-reversal symmetry in 2D electronic systems leads to new topological invariants.~\cite{km2}  While nonzero Chern numbers cannot be realized with time-reversal invariance, the zero-Chern-number class of systems gets subdivided into two pieces: ``ordinary'' insulators that do not in general have an edge state, and a ``quantum spin Hall effect'' or ``topological insulator'' where a bulk topological invariant forces an edge state.  The topological invariant is not an integer but rather a two-valued or $\mathbb{Z}_2$ invariant.

One idea that triggered this development was to consider two copies of the quantum Hall effect, one for spin-up electrons and one for spin-down, with opposite effective magnetic fields for the two spins.  This combination~\cite{murakamishi} is time-reversal invariant because acting with the time-reversal operator $T$ changes both the magnetic field direction and the spin.  Note that in a model such as this, $S_z$ is a conserved quantum number even though $SU(2)$ (spin-rotation invariance) is clearly broken, as up and down spins behave differently.  Heuristically, think of the spin-orbit coupling as arising from intra-atomic terms like ${\bf L} \cdot {\bf S}$, and consider specifically $L_z S_z$.  For an electron of fixed spin, this coupling to the orbital motion described by $L_z$ is just like the coupling in a constant magnetic field.  In the simplest case of a Chern number $+1$ state of up electrons and a Chern number $-1$ state of down electrons, the edge will have counterpropagating modes: e.g., up-spin moves clockwise along the edge and down-spin moves counterclockwise.  This turns out not to be a bad caricature of the quantum spin Hall phase in a more realistic system: one can tell by symmetry arguments that it will have no quantum Hall effect (i.e., $\alpha_c = 0$ in $J_i = \alpha_c \epsilon_{ijk} E_j B_k$), but it will generically have a ``spin Hall effect''
\beq
J^i_j = \alpha_s \epsilon_{ijk} E_k,
\eeq
where $\alpha_c$ and $\alpha_s$ are numbers and $J^i_j$ is a spin current (a current of angular momentum $i$ in spatial direction $j$).

\begin{figure}[!ht]
	\includegraphics[scale=0.4]{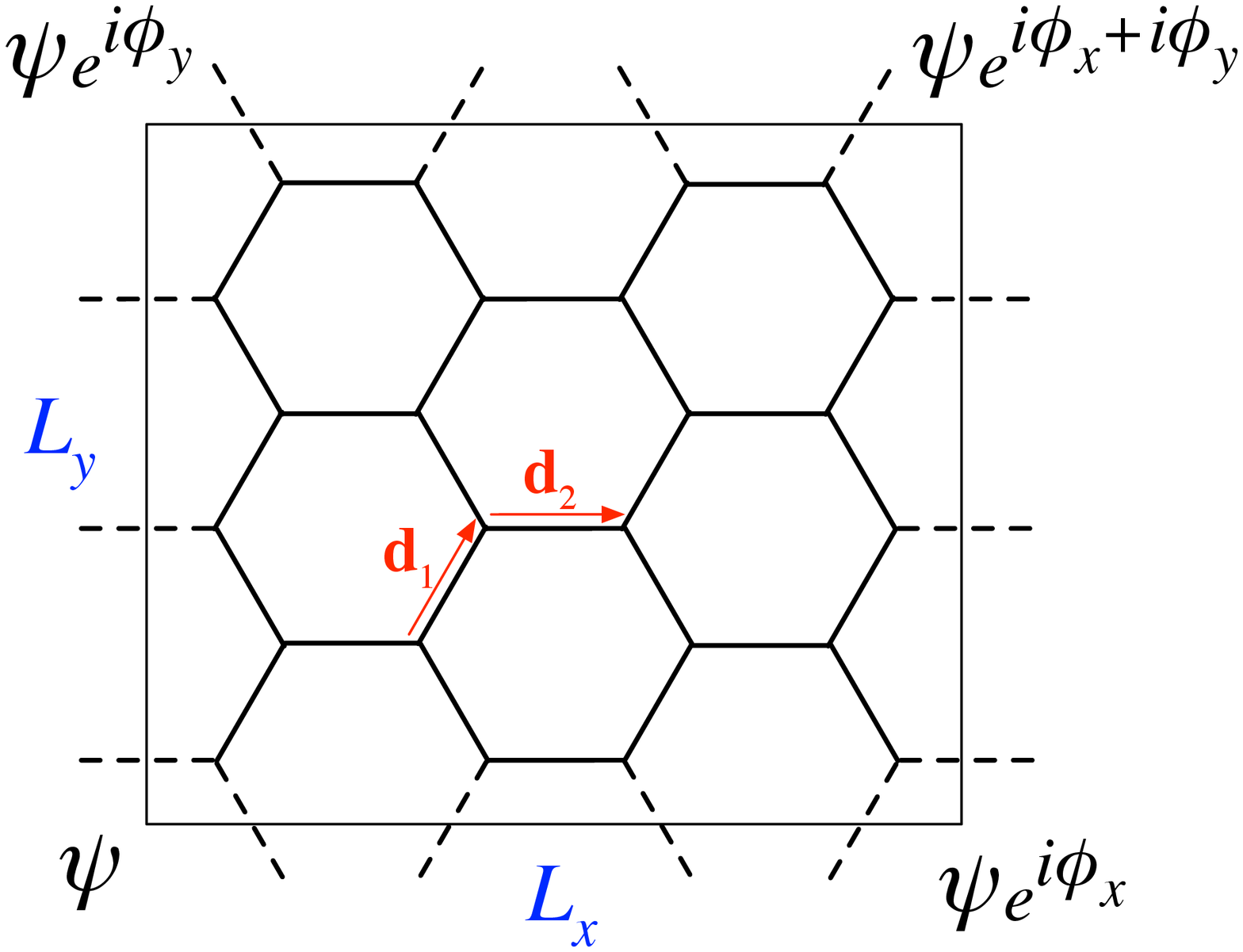}
	\caption{(Color online) The honeycomb lattice on which the tight-binding
	  model graphene Hamiltonian of Kane and Mele~\cite{kane&mele-2005} is defined.
		For the two sites depicted, the factor $\nu_{ij}$ of equation (\ref{HSO}) 
		is $\nu_{ij} = -1$.  The phases $\phi_{x,y}$ describe twisted boundary
		conditions that can be used to give a pumping definition of the $\mathbb{Z}_2$ invariant in the disordered case.  (From Essin and Moore~\cite{essinmoore}.)
	\label{honeycomb}}
\end{figure}

As an example of this ``two copies of the IQHE'' generated by spin-orbit coupling, consider the model of graphene introduced by Kane and Mele.\cite{kane&mele-2005}  This is a tight-binding model for independent electrons on the honeycomb lattice  (Fig.~\ref{honeycomb}).  The spin-independent part of the Hamiltonian consists of a nearest-neighbor hopping, which alone would give a semimetallic spectrum with Dirac nodes at certain points in the 2D Brillouin zone, plus a staggered sublattice potential whose effect is to introduce a gap:
\beq\label{H0}
H_0 = t \sum_{\langle i j \rangle \sigma} c^\dagger_{i \sigma} c_{j \sigma} + \lambda_v \sum_{i \sigma} \xi_i c^\dagger_{i \sigma} c_{i \sigma}.
\eeq
Here $\langle ij \rangle$ denotes nearest-neighbor pairs of sites, $\sigma$ is a spin index, $\xi_i$ alternates sign between sublattices of the honeycomb, and $t$ and $\lambda_v$ are parameters.

The insulator created by increasing $\lambda_v$ is an unremarkable band insulator.  However, the symmetries of graphene also permit an ``intrinsic'' spin-orbit coupling of the form
\beq\label{HSO}
H_{SO} = i \lambda_{SO} \sum_{\langle \langle ij \rangle \rangle \sigma_1 \sigma_2} \nu_{ij} c^\dagger_{i \sigma_1} s^z_{\sigma_1 \sigma_2} c_{j \sigma_2}.
\eeq
Here $\nu_{ij} = (2 / \sqrt{3}) \hat{\bm{d}}_1 \times \hat{\bm{d}}_2 = \pm 1$, where $i$ and $j$ are next-nearest-neighbors and $\hat{\bm{d}}_1$ and $\hat{\bm{d}}_2$ are unit vectors along the two bonds that connect $i$ to $j$.  Including this type of spin-orbit coupling alone would not be a realistic model.  For example, the Hamiltonian $H_0+H_{SO}$ conserves $s^z$, the distinguished component of electron spin, and reduces for fixed spin (up or down) to Haldane's model.\cite{haldane1988}  Generic spin-orbit coupling in solids should not conserve any component of electron spin.

This model with $S_z$ conservation is mathematically treatable using the Chern number above, as it just reduces to two copies of the IQHE.  It turns out that the stability of the phase to more general spin-orbit coupling is dependent on a subtle property of spin-half particles.  The surprise is that the quantum spin Hall phase survives, with interesting modifications, once we allow more realistic spin-orbit coupling, as long as time-reversal symmetry remains unbroken.  The time-reversal operator $T$ acts differently in Fermi and Bose systems, or more precisely in half-integer versus integer spin systems.  Kramers showed that the square of the time-reversal operator is connected to a $2 \pi$ rotation, which implies that
\beq
T^2 = (-1)^{2 S},
\eeq
where $S$ is the total spin quantum number of a state: half-integer-spin systems pick up a minus sign under two time-reversal operations.

An immediate consequence of this is the existence of ``Kramers pairs'': every eigenstate of a time-reversal-invariant spin-half system is at least two-fold degenerate.
Combining Kramers pairs with what is known about the edge state gives some intuition about why a odd-even or $\mathbb{Z}_2$ invariant might be physical here.  If there is only a single Kramers pair of edge states and we consider low-energy elastic scattering, then a right-moving excitation can only backscatter into its time-reversal conjugate, which is forbidden by the Kramers result (it would split the pair) if the perturbation inducing scattering is time-reversal invariant.  However, if we have two Kramers pairs of edge modes, then a right-mover can back-scatter to the left-mover that is {\it not} its time-reversal conjugate~\cite{xumooreedge,congjun}.  This process will, in general, eliminate these two Kramers pairs from the low-energy theory.

The general belief based on this argument is that a system with an even number of Kramers pairs will, under time-reversal-invariant backscattering, localize in pairs down to zero Kramers pairs, while a system with an odd number of Kramers pairs will wind up with a single stable Kramers pair.  But rather than just trying to understand whether the edge is stable, to predict from bulk properties whether the edge will have an even or odd number of Kramers pairs.  Kane and Mele showed that starting from the 2D $T$-invariant crystal, there are two topological classes; instead of an integer-valued topological invariant, there is a ``$\mathbb{Z}_2$ invariant with two values, ``even'' and ``odd''.  Several formulations of this invariant are now available~\cite{km2,bernevigzhangz2,royz2,fu&kane1-2006,moore&balents-2006}.  The even and odd classes correspond in the example above (of separated up- and down-spins) to paired IQHE states with even or odd Chern number for one spin.

We expect that, as for the IQHE, it should be possible to reinterpret the $\mathbb{Z}_2$ invariant as an invariant that describes the response of a finite toroidal system to some perturbation.  In the IQHE, the response is the amount of charge that is pumped around one circle of the torus as a $2 \pi$ flux (i.e., a flux $hc/e$) is pumped adiabatically through the other circle.  The pumping definition for the 2D topological insulator involves a $\pi$-flux; one can consider pumping of ``$\mathbb{Z}_2$'' (or Kramers degeneracy) from one boundary to another of a large cylinder~\cite{fu&kane1-2006} (local insertion of a $\pi$ flux similarly induces a degeneracy~\cite{ranvishwanathlee}).  Alternately, a precise restatement of  the $\mathbb{Z}_2$ invariant closes the process after $\pi$ flux insertion so that what is pumped is ordinary charge.~\cite{essinmoore,qipumping}  These approaches directly generalize the $\mathbb{Z}_2$ invariant to disordered single-electron systems; this and other 3D effects have recently been studied~\cite{shindou}.
For the 3D topological insulator, we will later introduce an alternative to pumping, the magnetoelectric polarizability, that defines the phase in the presence of interactions.

\subsection{3D topological insulators}

We review quickly the band structure invariants that were used in the first definitions of the 3D topological insulator~\cite{moore&balents-2006,fu&kane&mele-2007,rroy3D}.  Since the key emergent properties are somewhat difficult to perceive directly from the bulk band structure invariants, we then use a different picture to explain the surface state~\cite{fu&kane&mele-2007} and magnetoelectric polarizability~\cite{qilong,essinmoorevanderbilt,essinturnermoorevanderbilt}.

We start by asking to what extent the two-dimensional integer quantum Hall effect can be generalized to three dimensions.  A generalization of the previously mentioned homotopy argument of Avron, Seiler, and Simon~\cite{ass} can be used to show that there are three Chern numbers per band in three dimensions, associated with the $xy$, $yz$, and $xz$ planes of the Brillouin zone.  A more physical way to view this is that a three-dimensional integer quantum Hall system consists of a single Chern number and a reciprocal lattice vector that describes the ``stacking'' of integer quantum Hall layers~\cite{halperin3D}.  The edge of this three-dimensional IQHE is quite interesting: it can form a two-dimensional chiral metal, as the chiral modes from each IQHE~\cite{halperin} combine and point in the same direction.~\cite{chalkerchiral,fisherchiral}

Now consider the Brillouin zone of a three-dimensional time-reversal-invariant material.  Our approach, following the derivation of the 3D topological invariants in~\cite{moore&balents-2006}, builds on the two-dimensional case.  In the case of inversion symmetry, a useful trick~\cite{fu&kane2-2006,hasankane} allows the integrals in the following approach to be expressed simply in terms of the parity eigenvalues at the time-reversal-invariant points.  This led to the understanding of the detailed surface state structure~\cite{fu&kane&mele-2007} that was important the experimental observation of the phase~\cite{hsieh}.

Concentrating on a single band pair, there is a $\mathbb{Z}_2$ topological invariant defined in the two-dimensional problem with time-reversal invariance.  Taking the Brillouin zone to be a torus, there are two inequivalent $xy$ planes that are distinguished from others by the way time-reversal acts: the $k_z = 0$ and $k_z = \pm \pi / a$ planes are taken to themselves by time-reversal (note that $\pm \pi/a$ are equivalent because of the periodic boundary conditions).  These special planes are essentially copies of the two-dimensional problem, and we can label them by $\mathbb{Z}_2$ invariants $z_0 = \pm 1$, $z_{\pm 1} = \pm 1$, where $+1$ denotes ``even Chern parity'' or ordinary 2D insulator and $-1$ denotes ``odd Chern parity'' or topological 2D insulator.  Other $xy$ planes are not constrained by time-reversal and hence do not have to have a $\mathbb{Z}_2$ invariant.

The most interesting 3D topological insulator phase (the ``strong topological insulator'', in the language of~\cite{fu&kane&mele-2007}) results when the $z_0$ and $z_{\pm 1}$ planes are in different 2D classes.  This can occur if, moving in the $z$ direction between these two planes, one has a series of 2D problems that interpolate between ordinary and topological insulators by breaking time-reversal.  We will concentrate on this type of 3D topological insulator here.  Another way to make a 3D topological insulator is to stack 2D topological insulators, but considering the edge of such a system shows that it will not be very stable: since two ``odd'' edges combine to make an ``even'' edge, which is unstable in the presence of $T$-invariant backscattering, we call such a stacked system a ``weak topological insulator''~\cite{fu&kane&mele-2007}.  This edge state can appear at crystalline dislocations~\cite{ranzhangvishwanath}.

Above we found two $xy$ planes with two-dimensional $\mathbb{Z}_2$ invariants.  By the same logic, we could identify four other such invariants $x_0$, $x_{\pm 1}$, $y_0$, $y_{\pm 1}$.  However, not all six of these invariants are independent: some geometry shows that there are two relations, reducing the number of independent invariants to four:
\beq
x_0 x_{\pm 1} = y_0 y_{\pm 1} = z_0 z_{\pm 1}.
\eeq
(Sketch of geometry: to establish the first equality above, consider evaluating the Fu-Kane 2D formula on the four EBZs described by the four invariants $x_0, x_{+1}, y_0, y_{+1}$.  These define a torus, on whose interior the Chern two-form $F$ is well-defined.  Arranging the four invariants so that all have the same orientation, the $A$ terms drop out, and the $F$ integral vanishes as the torus can be shrunk to a loop.  In other words, for some gauge choice the difference $x_0 - x_{+1}$ is equal to $y_0 - y_{+1}$. )
We can take these four invariants in three dimensions as $(x_0,y_0,z_0,x_0 x_{\pm 1})$, where the first three describe layered ``weak'' topological insulators, and the last describes the strong topological insulator (STI), which is the one typically meant by "3D topological insulator".  The ``axion electrodynamics'' field theory in the next subsection is consistent with there being only one genuinely three-dimensional $\mathbb{Z}_2$ invariant.

This derivation shows that the strong topological insulator cannot be realized in any model with $S_z$ conservation, while, as explained earlier, a useful example of the 2D topological insulator (a.k.a. ``quantum spin Hall effect'') can be obtained from combining IQHE phases of up and down electrons.  The impossibility of making an STI with $S_z$ conservation follows from noting that all planes normal to $z$ have the same Chern number, as Chern number is a topological invariant whether or not the plane is preserved by time-reversal.  In particular, the $k_z= 0$ and $k_z = \pm \pi/a$ phases have the same Chern number for up electrons, say, which means that these two planes are either both 2D ordinary or 2D topological insulators.

While the above argument is rather general, it does not give much insight into what sort of gapless surface states we should expect at the surface of a strong topological insulator.  The spin-resolved surface Fermi surface encloses an odd number of Dirac points~\cite{fu&kane&mele-2007}.  In the simplest case of a single Dirac point, believed to be realized in Bi$_2$Se$_3$, the surface state can be pictured as ``one-quarter of graphene.''  Graphene, a single layer of carbon atoms that form a honeycomb lattice, has two Dirac points and two spin states at each $k$; spin-orbit coupling is quite weak since carbon is a relatively light element.  The surface state of a three-dimensional topological insulator can have a single Dirac point and a single spin state at each $k$.  As in the edge of the 2D topological insulator, time-reversal invariance implies that the spin state at $k$ must be the $T$ conjugate of the spin state at $-k$.  This picture of the surface state can be obtained by working backward; in the following section we will effectively redefine a topological insulator as a material whose surfaces, when gapped by a weak magnetic field, have half-integer quantum hall effect, and since each Dirac fermion at the surface contributes a half-integer when gapped, there must have been an odd number prior to the magnetic perturbation.

\section{TOPOLOGICAL INSULATORS AND ORBITAL MAGNETOELECTRIC POLARIZABILITY}

The three-dimensional topological insulator turns out to be connected to a basic electromagnetic property of solids: the orbital contribution to the magnetoelectric polarizability $dP/dB$~\cite{qilong,essinmoorevanderbilt,essinturnermoorevanderbilt}.  We know that in an insulating solid, Maxwell's equations can be modified because the dielectric constant $\epsilon$ and magnetic permeability $\mu$ need not take their vacuum values.  Another effect is that solids can generate the electromagnetic term
\beq
\Delta {\cal L}_{EM} = \frac{\theta e^2}{2\pi h} {\bf E} \cdot {\bf B} =
\frac{\theta e^2}{16\pi h} \epsilon^{\alpha \beta \gamma \delta}
	F_{\alpha \beta} F_{\gamma \delta}.
\label{axioncoupling}
\eeq
This term describes a magnetoelectric polarizability: an applied electrical field generates a magnetic dipole, and vice versa.  An essential feature of the above ``axion electrodynamics'' (cf. Wilczek, 1987~\cite{wilczekaxion} and references therein) is that, when the axion
field $\theta(\bm{x},t)$ is constant, it plays no role
in electrodynamics; this follows because $\theta$ couples
to a total derivative, $\epsilon^{\alpha \beta \gamma \delta}
F_{\alpha \beta} F_{\gamma \delta} = 2 \epsilon^{\alpha \beta
\gamma \delta} \partial_\alpha (A_\beta F_{\gamma \delta})$ (here we used that $F$ is closed, i.e., $dF=0$), and
so does not modify the equations of motion.  However, the presence
of the axion field can have profound consequences at surfaces and
interfaces, where gradients in $\theta(\bm{x})$ appear.

A bit of work shows that, at a surface where $\theta$ changes, there is a surface quantum Hall layer of magnitude
\beq
\sigma_{xy } = {e^2 (\Delta \theta) \over 2 \pi h}.
\eeq
(This can be obtained by moving the derivative from one of the $A$ fields to act on $\theta$, leading to a Chern-Simons term for the EM field at the surface, which gives an integer quantum Hall effect.)
The magnetoelectric polarizability described above can be obtained from these layers: for example, an applied electric field generates circulating surface currents, which in turn generate a magnetic dipole moment.  In a sense, $\sigma_{xy}$ is what accumulates at surfaces because of the magnetoelectric polarizability, in the same way as charge is what accumulates at surfaces because of ordinary polarization.

We write $\theta$ as an angle because, like polarization, $\theta$ is only well defined as a bulk property modulo $2 \pi$.  The integer multiple of $2 \pi$ is only specified once we specify a particular way to make the boundary.  The connection to the 3D topological insulator?~\cite{qilong}  At first glance, $\theta=0$ in any time-reversal-invariant system, since $\theta \rightarrow -\theta$ under time-reversal.  However, since $\theta$ is periodic, $\theta = \pi$ also works, as $-\theta$ and $\theta$ are equivalent because of the periodicity, and is inequivalent to $\theta = 0$.



In a band structure of noninteracting electrons, $\theta$ receives a contribution from the integral of the Chern-Simons form:
\beq
\theta=-\frac{1}{4\pi}\int_{\rm BZ} d^3k \; \epsilon_{ijk}\,
{\rm Tr}[\A_i\del_j\A_k-i\frac{2}{3}\A_i\A_j\A_k].
\label{theta}
\eeq
Qi, Hughes, and Zhang obtained this term by working down from the 4D quantum Hall effect, as explained below.  It can also be obtained~\cite{essinmoorevanderbilt} from semiclassical results on polarization in inhomogeneous systems~\cite{xiao} or more direct approaches~\cite{essinturnermoorevanderbilt,malashevich}.  The improvement is that these make clear that, for a general solid (not necessarily a topological insulator), the Chern-Simons integral is just one contribution to the electromagnetic $\theta$.  Here we sketch on understanding the physical and mathematical meaning of the Chern-Simons form that constitutes the integrand, chiefly by discussing analogies with our previous treatment of polarization in one dimension and the IQHE in two dimensions.  These analogies and some important differences are summarized in Table I.

In this section,
\beq
{\cal F}_{ij} = \partial_i {\cal A}_j - \partial_j
{\cal A}_i -i [{\cal A}_i, {\cal A}_j]
\eeq
is the (generally non-Abelian) Berry curvature tensor ($\A_\lambda=i\me{u}{\del_\lambda}{u}$),
and the trace and commutator refer to band indices.
We can understand the Chern-Simons form $K = {\rm Tr} [\A_i\del_j\A_k-i\frac{2}{3}\A_i\A_j\A_k]$ that appears in the integral above starting from the second Chern form ${\rm Tr} [\F \wedge \F]$; the relationship between the two is that
\beq
dK = {\rm Tr} [\F \wedge \F],
\eeq
just as $\A$ is related to the first Chern form: $d({\rm Tr} \A) = {\rm Tr} \F$.  These relationships hold locally (this is known as Poincare's lemma, that any closed form is {\it locally} an exact form) but not globally, unless the first or second Chern form is trivial (i.e., generates the trivial cohomology class.  For example, the existence of a nonzero first Chern number on the sphere prevents us from finding globally defined wavefunctions that would give an $\A$ with $d\A=\F$.  We are assuming in even writing the Chern-Simons formula for $\theta$ that the ordinary Chern numbers are zero, so that an $\A$ can be defined in the 3D Brillouin zone.  We would run into trouble if we assumed that an $\A$ could be defined in the 4D Brillouin zone if the {\it first or second} Chern number were nonzero.  
The second Chern number invariant of a 4D band structure~\cite{asss} underlies the four-dimensional quantum Hall effect~\cite{zhanghu}.


Although this 4D QHE is not directly measurable, it is mathematically connected to the 3D magnetoelectric polarizability in the same way as 1D polarization and the 2D IQHE are connected.  But a difference is that the Berry phase entirely determines 1D polarization, while in general the orbital magnetoelectric polarizability of a 3D crystal, even its diagonal part, contains further terms beyond the Berry phase (the correct expression is in~\cite{essinturnermoorevanderbilt,malashevich}).  But these ``ordinary''  terms disappear in either the ``topological'' limit of~\cite{qilong} where all occupied bands have the same energy -1 and all unoccupied bands have the same energy +1, or in topological insulators, where the only surviving term is (Eq.~\ref{theta}).  Hence in many simple models in the literature the extra terms actually vanish; an exception is a pyrochlore lattice model~\cite{guo}.

\begin{table}
\begin{tabular*}{4 in}{@{\extracolsep{\fill}}  c  c  c }
&Polarization&Magnetoelectric\cr
&&polarizability\cr
\hline\hline\\
$d_{\rm min}$&1&3\\
Observable & ${\bf P} = {\partial \langle H \rangle / \partial E}$ & $M_{ij} = \partial \langle H \rangle / \partial E_i \partial B_j$\cr
&&$= \delta_{ij} \theta e^2/(2 \pi h)$\cr
Quantum& $\Delta {\bf P} = e {\bf R}/ \Omega$ &$\Delta M = e^2/h$ \cr
Surface&$q=({\bf P}_1 - {\bf P}_2) \cdot {\bf \hat n}$&$\sigma_{xy} = (M_1 - M_2)$\cr
EM coupling & ${\bf P} \cdot {\bf E}$ & $M {\bf E} \cdot {\bf B} $\cr
Chern-Simons form&${\cal A}_i$&
$\epsilon_{ijk} ({\cal A}_i {\cal F}_{jk} + i {\cal A}_i {\cal A}_j {\cal A}_k/3)$\cr
Chern form& $\epsilon_{ij} \partial_i {\cal A}_j $&$ \epsilon_{ijkl} {\cal F}_{ij} {\cal F}_{kl}$\cr
Quantizing symmetry & Inversion & Inversion or ${\cal T}$\cr
Additional contributions? & No & Yes\cr
\end{tabular*}
\caption{Comparison of Berry-phase theories of polarization and magnetoelectric polarizability.}
\end{table}

%
%
%
%


We can give a many-body understanding of $\theta$ that clarifies the geometric reason for its periodicity even in a many-particle system~\cite{essinmoorevanderbilt}.  Consider evaluating $dP/dB$ by applying the 3D formula
\begin{equation} \label{polarization}
 P_i = e \int_{BZ} \! \frac{d^3 k}{(2\pi)^3} \mathrm{Tr}\,\mathcal{A}_i \,.
\end{equation}
which generalizes to interacting systems~\cite{ortizmartin} via the flux trick, to a rectangular-prism unit cell.  The minimum magnetic field normal to one of the faces that can be applied to the cell without destroying the periodicity is one flux quantum per unit cell~\cite{zak}, or a field strength $h/(e \Omega)$, where $\Omega$ is the area of that face.  The ambiguity of polarization (\ref{polarization}) in this direction is one charge per transverse unit cell area, i.e., $e/\Omega$.  Then the ambiguity in $dP/dB$ is, choosing the $x$ direction,
\begin{equation}
 \Delta \frac{P_x}{B_x} = {e / \Omega \over h / (e \Omega)} = \frac{e^2}{h} = 2\pi \frac{e^2}{2\pi h}.
 \label{quantumratio}
\end{equation}
So the periodicity of $2 \pi$ in $\theta$ is a consequence of the geometry of polarization, and is independent of the single-electron assumption that leads to the microscopic Chern-Simons formula.  The surface interpretation of the periodicity is that it results from the freedom to add a surface integer quantum Hall layer without any change in the bulk properties.  (Adding a fractional quantum Hall layer involves additional subtleties related to the topological ground state degeneracy.)

\section{RECENT DEVELOPMENTS: STRONG CORRELATIONS AND MAJORANA FERMIONS}

Many of the topics of greatest recent interest, both theoretically and experimentally, involve connections between strongly correlated electrons and topological insulators.  There could be materials in which the topological insulator behavior emerges as a result of electron correlations (a 2D theoretical model is introduced in~\cite{raghu}).  Similar behavior to that in simple topological insulators can occur in Kondo lattice systems~\cite{kondotopological}) and in antiferromagnets~\cite{mongessinmoore}.  In Mott insulators, emergent fermionic excitations (spinons) could form a topological phase with surfaces that are electrically insulating but thermally conducting~\cite{pesinbalents}.  Interactions could lead to spontaneous instabilities at the surfaces of topological insulators~\cite{xusurface}.

It seems fitting to end our review by explaining how topological insulators might enable creation and observation of one of the most long-sought particles of nature.  A Majorana fermion is a fermion that is its own antiparticle, unlike an ordinary (Dirac) fermion such as the electron, which has a distinct antiparticle (positron or hole).  The search for Majorana fermions as fundamental particles continues, but it is now believed that several condensed matter systems may support emergent Majorana fermion excitations, and possible observation of a Majorana fermion was reported in 2009 by Willett et al.~\cite{willettmajorana} by interferometric transport on the $\nu=5/2$ fractional quantum Hall state.

\begin{figure}[!ht]
	\includegraphics[scale=1.0]{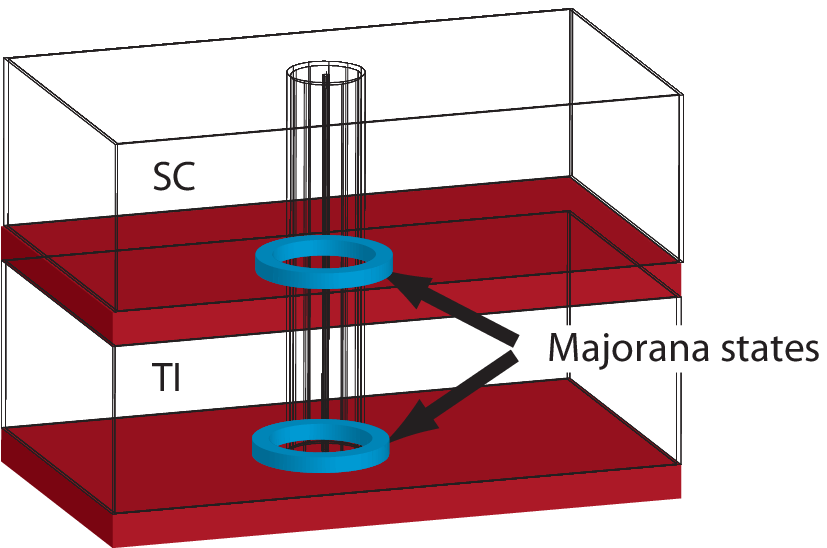}
	\caption{(Color online) The creation of Majorana fermions in the layer of unconventional superconductivity between an ordinary s-wave superconductor and a topological insulator~\cite{fukaneprox3d}.  A vortex core traps one Majorana fermion mode in its vicinity at each interface.
	\label{majorana}}
\end{figure}

Majorana fermions can be created in several ways using topological insulators.  The most direct proposal using a 3D topological insulator is to consider the proximity effect from an ordinary $s$-wave superconductor.  The result~\cite{fukaneprox3d} (Figure~\ref{majorana}) is an interface layer of 2D ``topological superconductor'' that is distinct but similar to the 2D $p+ip$ superconductor that has been understood for some time to support Majorana fermions.  A magnetic vortex core in such a system will carry a zero-energy Majorana fermion state localized near the vortex in the interface layer, as well as possibly ordinary electronic modes.

There have been many recent works discussing either how the Majorana mode could be detected (using, e.g., noise~\cite{nilsson}, interferometry~\cite{beenakkerinterfero,kaneinterfero}) and on collective behavior of Majoranas, e.g., in a random system of vortices~\cite{shivamoggi} or an array~\cite{xufu}).  There are analogous ways to create a Majorana fermion using strong spin-orbit quantum wells rather than topological insulators~\cite{fujimotosato,sau,aliceamajorana}.  These ideas for probes may also be helpful in understanding whether Cu$_x$Bi$_2$Se$_3$~\cite{cavasc,fuberg} is a topological superconductor.

We have tried to convey that, only two years after the first experimental paper on topological insulators, work in this field is proceeding on many fronts.  Our basic ideas of such established concepts as spin-orbit coupling and band structure have already had to be rethought, and there are good reasons to anticipate further advances in our understanding.  The authors have enjoyed too many collaborations and discussions in this field to acknowledge all by name; they are especially grateful to their colleagues for creating stimulating academic environments at Princeton (MZH) and Berkeley (JEM).  Funding was provided by DOE and NSF (MZH) and by NSF DMR-0804413 and WIN (JEM).

\bibliographystyle{arnuke_revised}

\begin{thebibliography}{10}

\bibitem{hsieh}
Hsieh D, Qian D, Wray L, Xia Y, Hor YS, et~al.,
\newblock Nature 452:970 (2008).

\bibitem{hasankane}
Hasan MZ, Kane CL,
\newblock Rev. Mod. Phys. 82:3045 (2010).

\bibitem{moorenature}
Moore JE,
\newblock Nature 464:194 (2010).

\bibitem{moore&balents-2006}
Moore JE, Balents L,
\newblock Phys. Rev. B 75:121306(R) (2007).

\bibitem{kane&mele2-2005}
Kane CL, Mele EJ,
\newblock Phys. Rev. Lett. 95:146802 (2005).

\bibitem{fu&kane&mele-2007}
Fu L, Kane CL, Mele EJ,
\newblock Phys. Rev. Lett. 98:106803 (2007).

\bibitem{fu&kane2-2006}
Fu L, Kane CL,
\newblock Phys. Rev. B 76:045302 (2007).

\bibitem{zhangscience1}
{Bernevig} BA, {Hughes} TL, {Zhang} SC,
\newblock Science 314:1757 (2006).

\bibitem{molenkampscience}
{Koenig} M, {Wiedmann} S, {Bruene} C, {Roth} A, {Buhmann} H, et~al.,
\newblock Science 318:766 (2007).

\bibitem{qizhang}
Qi XL, Zhang SC,
\newblock Physics Today 63:33 (2010).

\bibitem{qshereview}
{K{\"o}nig} M, {Buhmann} H, {Molenkamp} LW, {Hughes} T, {Liu} C, et~al.,
\newblock Journal of the Physical Society of Japan 77:031007 (2008), 0801.0901.

\bibitem{schnyder}
Schnyder A, Ryu S, Furusaki A, Ludwig AWW,
\newblock Phys. Rev. B 78:195125 (2008).

\bibitem{kitaev}
Kitaev A,
\newblock Periodic table for topological insulators and superconductors,
\newblock arXiv:0901.0686, 2009.

\bibitem{rroy3D}
Roy R,
\newblock Phys. Rev. B 79:195322 (2009).

\bibitem{teo09}
Teo JCY, Fu L, Kane C,
\newblock Phys. Rev. B 78:045426 (2008).

\bibitem{lenoir}
Lenoir B, et~al.,
\newblock in Fifteenth International Conference on Thermoelectrics (Pasadena,
  California) 1Ð13 (IEEE, New York), 1996.

\bibitem{liuallen}
Liu Y, Allen E,
\newblock Phys. Rev. B 52:1566 (1995).

\bibitem{wolff}
Wolff PAJ,
\newblock Phys. Chem. Solids 25:1057 (1964).

\bibitem{hsieh09a}
Hsieh D, et~al.,
\newblock Science 323:919 (2009).

\bibitem{roushan}
Roushan P, et~al.,
\newblock Nature 460:1106 (2009).

\bibitem{xia09a}
Xia Y, et~al.,
\newblock Nature Phys. 5:398 (2009).

\bibitem{xiaarxiv}
Xia Y, et~al.,
\newblock arXiv:0812.2078v1, 2008.

\bibitem{hsieh09b}
Hsieh D, et~al.,
\newblock Nature 460:1101 (2009).

\bibitem{anderson}
Anderson PW,
\newblock Phys.\ Rev. 109:1492 (1958).

\bibitem{tzhang}
Zhang T, et~al.,
\newblock Phys. Rev. Lett. 103:266803 (2009).

\bibitem{alpichshev}
Alpichshev Z, et~al.,
\newblock Physical Review Letters 104:016401 (2010).

\bibitem{zhangn}
Zhang H, et~al.,
\newblock Nature Physics 5:438 (2009).

\bibitem{noh}
Noh HJ, et~al.,
\newblock Europhys. Lett. 81:57006 (2008).

\bibitem{hor09}
Hor YS, et~al.,
\newblock Phys. Rev. B 79:195208 (2009).

\bibitem{park10}
Park SR, et~al.,
\newblock Phys. Rev. B 81:041405 (2010).

\bibitem{chen}
Chen YL, et~al.,
\newblock Science 325:178 (2009).

\bibitem{hsieh09c}
Hsieh D, et~al.,
\newblock Phys. Rev. Lett. 103:146401 (2009).

\bibitem{xia09b}
Xia Y, et~al.,
\newblock arXiv:0907.3089, 2009.

\bibitem{joelrev}
Moore J,
\newblock Nature Physics 5:378 (2009).

\bibitem{lin2010}
Lin, et~al.,
\newblock Nature Materials 9:546 (2010).

\bibitem{fuwarping}
Fu L,
\newblock Phys. Rev. Lett. 103:266801 (2009).

\bibitem{hasan09}
Hasan MZ, Lin H, Bansil A,
\newblock Physics 2:108 (2009).

\bibitem{wray10}
Wray L, et~al.,
\newblock unpublished, 2010.

\bibitem{cavasc}
Hor YS, et~al.,
\newblock Phys. Rev. Lett. 104:057001 (2010).

\bibitem{wray09}
Wray L, et~al.,
\newblock arXiv:0912.3341v1, 2009.

\bibitem{chadov10}
Chadov, et~al.,
\newblock Nature Materials 9:541 (2010).

\bibitem{gapexp}
Li YY, et~al.,
\newblock Growth dynamics and thickness-dependent electronic structure of
  topological insulator {B}i$_2${T}e$_3$ thin film on {S}i,
\newblock arXiv:0912.5054, 2009.

\bibitem{cui}
Peng H, Lai K, Kong D, Meister S, Chen Y, et~al.,
\newblock Nat.\ Mat. 9:225 (2010).

\bibitem{bardarson}
Bardarson JH, Brouwer PW, Moore JE,
\newblock Aharonov-bohm oscillations in disordered topological insulator
  nanowires,
\newblock arXiv:1005.3762, 2010.

\bibitem{zhangvishwanath}
Zhang Y, Vishwanath A,
\newblock Anomalous aharonov-bohm conductance oscillations from topological
  insulator surface states,
\newblock arXiv:1005.3542, 2010.

\bibitem{gapp}
Lu HZ, et~al.,
\newblock Phys. Rev. B 37:115407 (2010).

\bibitem{franzmoore}
Seradjeh B, Moore JE, Franz M,
\newblock Phys. Rev. Lett. 103:066402 (2009).

\bibitem{tknn}
Thouless DJ, Kohmoto M, Nightingale MP, den Nijs M,
\newblock Phys. Rev. Lett. 49:405 (1982).

\bibitem{berry}
Berry MV,
\newblock Proc.\ Roy.\ Soc.\ A 392:45 (1984).

\bibitem{ksv}
King-Smith RD, Vanderbilt D,
\newblock Phys. Rev. B 47:1651 (1993).

\bibitem{nakahara}
Nakahara M,
\newblock {\em {Geometry, Topology and Physics}} (Institute of Physics, 1998).

\bibitem{resta}
Resta R,
\newblock Ferroelectrics 136:51 (1992).

\bibitem{asss}
Avron JE, Sadun L, Segert J, Simon B,
\newblock Phys. Rev. Lett. 61:1329 (1988).

\bibitem{zhanghu}
Zhang SC, Hu J,
\newblock Science 294:823 (2001).

\bibitem{niuthouless}
Niu Q, Thouless DJ, Wu YS,
\newblock Phys. Rev. B 31:3372 (1985).

\bibitem{ass}
Avron JE, Seiler R, Simon B,
\newblock Phys. Rev. Lett. 51:51 (1983).

\bibitem{km2}
Kane CL, Mele EJ,
\newblock Phys. Rev. Lett. 95:226801 (2005).

\bibitem{murakamishi}
Murakami S, Nagaosa N, Zhang SC,
\newblock Phys. Rev. Lett. 93:156804 (2004).

\bibitem{kane&mele-2005}
Kane CL, Mele EJ,
\newblock Phys. Rev. Lett. 95:146802 (2005).

\bibitem{haldane1988}
{Haldane} FDM,
\newblock Physical Review Letters 61:2015 (1988).

\bibitem{essinmoore}
Essin AM, Moore JE,
\newblock Phys. Rev. B 76:165307 (2007).

\bibitem{xumooreedge}
Xu C, Moore JE,
\newblock Phys. Rev. B 73:045322 (2006).

\bibitem{congjun}
Wu C, Bernevig BA, Zhang SC,
\newblock Phys. Rev. Lett. 96:106401 (2006).

\bibitem{bernevigzhangz2}
Bernevig BA, Zhang SC,
\newblock Phys. Rev. Lett. 96:106802 (2006).

\bibitem{royz2}
Roy R,
\newblock Phys. Rev. B 79:195321 (2009).

\bibitem{fu&kane1-2006}
Fu L, Kane CL,
\newblock Phys. Rev. B 74:195312 (2006).

\bibitem{ranvishwanathlee}
Ran Y, Vishwanath A, Lee DH,
\newblock Phys. Rev. Lett. 101:086801 (2008).

\bibitem{qipumping}
Qi XL, Zhang SC,
\newblock Phys. Rev. Lett. 101:086802 (2008).

\bibitem{shindou}
Shindou R, Murakami S,
\newblock Phys. Rev. B 79:045321 (2009).

\bibitem{qilong}
Qi XL, Hughes TL, Zhang SC,
\newblock Physical Review B 78:195424 (2008).

\bibitem{essinmoorevanderbilt}
Essin AM, Moore JE, Vanderbilt D,
\newblock Physical Review Letters 102:146805 (2009).

\bibitem{essinturnermoorevanderbilt}
Essin AM, Turner AM, Moore JE, Vanderbilt D,
\newblock Physical Review B 81:205104 (2010).

\bibitem{halperin3D}
Halperin BI,
\newblock Jpn. J. Appl. Phys. Suppl. 26:1913 (1987).

\bibitem{halperin}
Halperin BI,
\newblock Phys. Rev. B 25:2185 (1982).

\bibitem{chalkerchiral}
Chalker JT, Dohmen A,
\newblock Phys. Rev. Lett. 75:4496 (1995).

\bibitem{fisherchiral}
Balents L, Fisher MPA,
\newblock Phys. Rev. Lett. 76:2782 (1996).

\bibitem{ranzhangvishwanath}
Ran Y, Zhang Y, Vishwanath A,
\newblock Nature Phys. 5:298 (2009).

\bibitem{wilczekaxion}
Wilczek F,
\newblock Phys. Rev. Lett. 58:1799 (1987).

\bibitem{xiao}
Xiao D, Shi J, Clougherty DP, Niu Q,
\newblock Phys. Rev. Lett. 102:087602 (2009).

\bibitem{malashevich}
Malashevich A, Souza I, Coh S, Vanderbilt D,
\newblock New Journal of Physics 12:053032 (2010).

\bibitem{guo}
Guo HM, Franz M,
\newblock Physical Review Letters 103:206805 (2009).

\bibitem{ortizmartin}
Ortiz G, Martin RM,
\newblock Phys. Rev. B 49:14202 (1994).

\bibitem{zak}
Zak J,
\newblock Phys. Rev. 134:A1602 (1964).

\bibitem{raghu}
Raghu S, Qi XL, Honerkamp C, Zhang SC,
\newblock Phys. Rev. Lett. 100:156401 (2008).

\bibitem{kondotopological}
Dzero M, Sun K, Galitski V, Coleman P,
\newblock Phys. Rev. Lett. 104:106408 (2010).

\bibitem{mongessinmoore}
Mong R, Essin AM, Moore JE,
\newblock Phys. Rev. B 81:245209 (2010).

\bibitem{pesinbalents}
Pesin DA, Balents L,
\newblock Nature Physics 6:376 (2010).

\bibitem{xusurface}
Xu C,
\newblock Phys. Rev. B 81:054403 (2010).

\bibitem{willettmajorana}
Willett RL, Pfeiffer LN, West KW,
\newblock PNAS 106:8853 (2009).

\bibitem{fukaneprox3d}
Fu L, Kane CL,
\newblock Phys. Rev. Lett. 100:096407 (2008).

\bibitem{nilsson}
Nilsson J, Akhmerov AR, Beenakker CW,
\newblock Phys. Rev. Lett. 101:120403 (2008).

\bibitem{beenakkerinterfero}
Akhmerov AR, Nilsson J, Beenakker CWJ,
\newblock Phys. Rev. Lett. 102:216404 (2009).

\bibitem{kaneinterfero}
Fu L, Kane CL,
\newblock Phys. Rev. Lett. 102:216403 (2009).

\bibitem{shivamoggi}
Shivamoggi V, Refael G, Moore JE,
\newblock Majorana fermion chain at the quantum spin hall edge,
\newblock arXiv:1004.4585, 2010.

\bibitem{xufu}
Xu C, Fu L,
\newblock Phys. Rev. B 81:134435 (2010).

\bibitem{fujimotosato}
Sato M, Fujimoto S,
\newblock Phys. Rev. B 79:094504 (2009).

\bibitem{sau}
Sau JD, Lutchyn RM, Tewari S, Sarma SD,
\newblock Phys. Rev. Lett. 104:040502 (2010).

\bibitem{aliceamajorana}
Alicea J,
\newblock Phys. Rev. B 81:125318 (2010).

\bibitem{fuberg}
Fu L, Berg E,
\newblock arXiv:0912.3294, 2010.

\end{thebibliography}

\listoffigures
\end{document}